\documentclass[12pt,preprint]{aastex}

\shorttitle{Energy release near the PIL}
\shortauthors{Sharykin et al.}

\begin{document}

\title{Flare Energy Release at the Magnetic Field Polarity Inversion Line During M1.2 Solar Flare of 2015 March 15. II. Investigation of Photospheric Electric Current and Magnetic Field Variations Using HMI 135-second Vector Magnetograms}

\author{I.N. Sharykin\altaffilmark{1,2}, I.V. Zimovets\altaffilmark{1}, I.I.~Myshyakov\altaffilmark{3}}

\affil{Space Research Institute of Russian Academy of Sciences (IKI), Moscow, Russia}

\altaffiltext{1}{Space Research Institute (IKI) of the Russian Academy of Sciences}
\altaffiltext{2}{Moscow Institute of Physics and Technology}
\altaffiltext{3}{Institute of Solar-Terrestrial Physics (ISTP) of the Russian Academy of Sciences, Siberian Branch}

%%%%%%%%%%%%%%%%%%%%%%%%%%%%%%%%%%%%%%%%%%%%%%%%%%%%%%%%%%%%%%%%%%%%%%%%%%%%%%%%%%%%%%%%%%%%%%%%%%%%%%%%%%%%%

\begin{abstract}

This work is a continuation of Paper~I \citep{Sharykin2018} devoted to analysis of nonthermal electron dynamics and plasma heating in the confined M1.2 class solar flare SOL2015-03-15T22:43 revealing energy release in the highly sheared interacting magnetic loops in the low corona, above the polarity inversion line (PIL). The scope of the present work is to make the first extensive quantitative analysis of the photospheric magnetic field and photospheric vertical electric current (PVEC) dynamics in the confined flare region near the PIL using new vector magnetograms obtained with the Helioseismic and Magnetic Imager (HMI) onboard the Solar Dynamics Observatory (SDO) with high temporal resolution of 135~s. Data analysis revealed sharp changes of the magnetic structure and PVEC associated with the flare onset near the PIL. It was found that the strongest plasma heating and electron acceleration were associated with the largest increase of the magnetic reconnection rate, total PVEC and effective PVEC density in the flare ribbons. Observations and non-linear force-free field (NLFFF) extrapolations showed that the magnetic field structure around the PIL is consistent with the tether-cutting magnetic reconnection (TCMR) geometry. We gave qualitative interpretation of the observed dynamics of the flare ribbons, magnetic field and PVEC, and electron acceleration, within the TCMR scenario.

\end{abstract}

%%%%%%%%%%%%%%%%%%%%%%%%%%%%%%%%%%%%%%%%%%%%%%%%%%%%%%%%%%%%%%%%%%%%%%%%%%%%%%%%%%%%%%%%%%%%%%%%%%%%%%%%%%%%%%%%%%%%%%%%%%%%%%%%%%%%%%%%%%%%%%%%%%%%%%%%%%%%%%%%%%%%%%
\section{INTRODUCTION}

Initiation of solar flares is connected with magnetic field dynamics in active regions of the Sun. Free magnetic energy contained in active regions in the form of electrical currents is enough to explain any small and large, confined and eruptive solar flares \citep{Emslie2012,Aschwanden2014}. One of the key observational objects determining magnetic field topology of active regions is the photospheric polarity inversion line (PIL). It has been known since the observations of \citet{Severnyi1958} that solar flares appear near the PIL of the line-of-sight (LOS) magnetic field \citep[e.g.][]{Rust1972,Schrijver2009,Sadykov2017}. The recent statistical study by \citet{Schrijver2016} illustrated that X-class flares are associated with strong-field, high-gradient polarity inversion lines (SHIL) created during the emergence of magnetic flux. Many recent works \citep[e.g.][]{Chifor2007,Zimovets2009a,Bamba2017a,Wang2017} also reported pre-flare activity around the PIL in the form of brightenings seen in different ranges of the electromagnetic spectrum. Thus, observations of the PIL regions are crucial for understanding the processes of accumulation and release of energy in solar flares.

Strong changes of photospheric magnetic field near the PIL during flares were reported in many works \citep[e.g.][]{Wang1994,Kosovichev2001,Sudol2005,Petrie2013,Fainshtein2016,Wang2018}. In particular, according to \citet{Sun2012,Petrie2013} magnetic field near the PIL became stronger and more horizontal, and the magnetic shear increased during flares. It was also found that the horizontal gradient of radial magnetic field increases during a flare \citep[e.g.][]{Sharykin2017}. The nonlinear force-free field (NLFFF) extrapolations have shown that magnetic field lines around the PIL became shorter during flares \citep[e.g.][]{Sun2012,Sharykin2015}. Probably, such behavior of magnetic loops can be connected with the process of magnetic reconnection above the photospheric PIL during flares.

According to the main modern paradigm, the primary energy release of solar flares, accompanied by the transformation of free magnetic energy into the kinetic energy of heated plasma and accelerated particles and radiation, occurs in the corona as a result of magnetic reconnection above the photospheric PIL \citep{Priest2002,Aschwanden2004,Shibata2011}. In a common situation, where there is one main PIL in an active region, a flare occurs in a bipolar magnetic configuration. Usually this bipolar structure is an arcade of sheared magnetic loops or a twisted magnetic flux rope embedded in the arcade along the PIL. Conventionally, the innermost part of the magnetic arcade is called the core field, and the outer layers are called the envelope fields \citep{Moore1992,Moore2001}. The core field lines are highly sheared, such they can almost be parallel to the PIL. The core field lines can constitute a magnetic flux-rope before a flare, or form it as a result of the magnetic reconnection during a flare \citep[e.g][]{Gibson2004,Gibson2006,Wang2015,Cheng2017}. The reconnection can also result in the increase of magnetic flux and twist of a pre-existing flux-rope.

The flare and eruption can be triggered in various ways in the bipolar magnetic configuration \citep[e.g.][]{Priest2014}. In particular, the widely discussed possibilities include (but not limited by) various instabilities of a magnetic flux-rope \citep[e.g.][]{Hood1979,Gibson2004,Torok2005,Kliem2006}, or the so-called tether-cutting magnetic reconnection (TCMR) of the core field leading to formation of an unstable flux-rope \citep{Moore1992,Moore2001}. The TCMR concept considers the situation when a group of highly sheared field lines interact and reconnect above the PIL, forming two new groups of field lines: one group of lower-lying, shorter and less sheared loops than initial field lines, and another group of higher field lines forming (or contributing to) a magnetic flux-rope. In such a scenario, a flare may develop in two consecutive or partially overlapping phases. The first phase is related to the TCMR or the so-called ``zipper-reconnection'' \citep{Priest2017,Threlfall2018a}, and is associated with the apparent systematic motion of $\mathrm{H_{\alpha}}$, ultraviolet (UV), soft X-ray (SXR; $\lesssim 10$ keV) and hard X-ray (HXR; $\gtrsim 10$ keV) emission sources along the PIL, and the elongation of flare ribbons \citep[e.g.][]{Vorpahl1976,Krucker2003,Bogachev2005,Grigis2005,Qiu2009,Yang2009,Zimovets2009b,Kuznetsov2016,Qiu2017}. This first phase usually coincides with the flare impulsive phase, often accompanied by successive bursts or pulsations of energetic electromagnetic radiation. The second phase is related to the formation of a three-dimensional quasi-vertical current sheet beneath an erupting flux-rope and magnetic reconnection in this current sheet. This phase is predominantly accompanied by expansion and separation of flare ribbons away from the PIL, and is well-explained by the ``standard'' (CSHKP) two-dimensional flare model and its three-dimensional extensions \citep{Carmichael1964,Sturrock1966,Hirayama1974,Kopp1976,Shibata2011,Janvier2017}.

One may suggest that depending on the initial configuration of the magnetic field and the development of the process of formation and eruption of a flux-rope, the two phases under consideration can be expressed differently. For example, in the case of the presence of a highly twisted unstable flux-rope before the onset of eruption and its quasi-homogeneous (symmetric) eruption along the PIL, the ``zipper-reconnection'' phase may be very quick or absent, and the motion along the PIL is not pronounced. In the event when a flux-rope eruption cannot fully develop, i.g., due to a strongly suppressing overlying magnetic field, the second phase of flare ribbon expansion may be mild or absent, and one observes a confined flare \citep[e.g.][]{Thalmann2015,Amari2018}. It should be noted here that the observed speeds of movement of the flare footpoint sources are used to estimate the reconnection electric field ($E$) and dimensionless reconnection rate in the coronal energy release sources, i.e. the Alfven Mach number $M_{A} = v_{in}/v_{A}$, where $v_{in}$ and $v_{A}$ are the inflow and Alfven speeds respectively \citep[e.g.][]{Forbes2000,Qiu2002,Isobe2002,Asai2004a,Isobe2005,Miklenic2007,Yang2011,Hinterreiter2018}. The values found in this way are in the ranges of $E \sim 1-100$~V~cm$^{-1}$ and $M_{A} \sim 0.001-0.1$.

Since the free energy in the active regions is contained in the form of electric currents flowing along magnetic field lines or having the form of current sheets, currents play an important role in the processes of flare energy release \citep{Melrose2017,Fleishman2018,Schmieder2018}. However, routine measurements of electric currents in the corona are not available yet. Measurements of the photospheric magnetic field vector in the active regions producing solar flares reveal sites of enhanced vertical electric currents in the vicinity of the PIL \citep[e.g.][]{Moreton1968,Leka1993,VanDrielGesztelyi1994,Janvier2014,Sharykin2015,Janvier2016,Wang2018}. Using vector magnetograms with high spatial resolution (up to 100~km) from the New Solar Telescope (NST) of the Big Bear Solar Observatory (BBSO) \citet{Wang2017} have shown that pre-flare activity in the form of optical brightenings near the PIL is associated with multiple inversions of the radial magnetic field and regions of enhanced photospheric vertical electric currents (PVEC).

One can suspect that electrons accelerated in the sheared magnetic structures, possibly associated with magnetic flux-ropes and/or the TCMR, may correspond to the regions of enhanced PVEC, as they trace twisted magnetic field lines stretched along the PIL. Many observations \citep[e.g.][]{Romanov1990,Abramenko1991,Canfield1992,Leka1993,Demoulin1996} have shown that flare $\mathrm{H_{\alpha}}$ or HXR emission sources do not exactly correspond to the most intensive PVEC. In particular, it was demonstrated by \citet{Li1997} that the sites of accelerated electrons precipitation to the chromosphere, found from the observations of the Hard X-ray Telescope (HXT) onboard Yohkoh, avoid the sites of the highest PVEC density and occur adjacent to these current channels. The recent studies by \citet{Musset2015,Sharykin2015a} using vector magnetograms from the Helioseismic and Magnetic Imager (HMI; \cite{Scherrer2012}) onboard the Solar Dynamics Observatory (SDO; \cite{Pesnell2012}) have shown that only a part of the HXR emission sources are located in the vicinity of the enhanced PVEC near the PIL. This indicates the absence of a direct spatial connection between the enhanced vertical electric currents and beams of accelerated particles. \citet{Musset2015} have also demonstrated co-spatial appearance of a new HXR source and a region of new enhanced PVEC during the same period of the powerful eruptive X2.2 flare on 15 February 2011. Moreover, it was shown recently that the photospheric vertical electric currents integrated over the flare regions near the PIL has tendency to increase abruptly and stepwise during some flares \citep{Janvier2014,Sharykin2015,Janvier2016,Wang2018}. These observations are not consistent with the electric circuit type models \citep{Alfven1967,Colgate1978,Zaitsev2015,Zaitsev2016}, where nonthermal electrons are assumed to be localized in magnetic loops with the strongest electric current, and the latter must dissipate, at least in part, during flares. Rather they support the concept that the processes of flare energy release and acceleration of particles occur in the coronal reconnection regions above the photospheric PIL. In this case, the presence of regions of enhanced PVEC indicates the presence of free magnetic energy above the PIL, a part of which can be released during a flare. The detected increase of vertical currents and the appearance of new sites of enhanced currents on the photosphere during flares can be interpreted as the generation of new currents in the coronal sources and their closure (at least partial) to or through the photosphere.

The major fraction of recent (since 2010) works dedicated to studying magnetic field and PVEC during flares were made using HMI/SDO vector magnetograms \citep{Centeno2014,Hoeksema2014}. Angular resolution of standard HMI vector magnetograms is about $1^{\prime\prime}$ when temporal resolution is 720~s. Comparing with typical duration of the flare impulsive phase ($\approx 1-10$~min), such time cadence is insufficient to resolve changes of magnetic field during flares. In this situation, one is mainly able to compare pre-flare and post-flare states of magnetic field. However, it is known that electron acceleration and plasma heating develops on shorter than 12~min time-scales \citep[e.g.][]{Aschwanden2004}. Thus, to understand development of the flare energy release processes one need to investigate dynamics of the flare emission sources, photospheric magnetic field and PVEC with a time resolution significantly less than 12~min.

New high-cadence 135-s vector magnetograms from HMI, which recently became in open access, have a large potential for this kind of research. Some long-duration ($\geq 10$ min) flare impulsive phases can be roughly investigated using these magnetograms. In the case of the shorter impulsive phases (with duration less than 2-3 min), it can be used to obtain more reliable change rates of magnetic flux and PVEC, because standard magnetograms give underestimated results due to averaging over the longer time interval of 12 min.

Standard (12-minute) HMI vector magnetograms are the result of summation of 135-second vector magnetograms. The time step of 135~s is an instrumental time needed for measuring the full set of Stokes profiles for calculation of all magnetic field components. Summation of 135-second magnetograms is made for increasing the signal-to-noise ratio. \citet{Sun2017} demonstrated that 135-second magnetograms are more noisy than the standard ones. However, for the magnetic field values larger than 300~G the uncertainty for all magnetic field components is small. \citet{Sun2017} described the high-cadence observations of magnetic field changes during the powerful X2.2 flare of 2011 February 15, which were mostly pronounced for the horizontal magnetic component. \citet{Kleint2018} presented the non-linear force-free field (NLFFF) modeling of the X1 flare region of 2014 March 29 based on 135-second magnetograms. They found that the magnetic field changes on the photosphere and in the chromosphere are surprisingly different, and are unlikely to be reproduced by a force-free model. These works demonstrated that the new HMI data product can be successfully used to study the flare processes in the active regions with strong magnetic fields. However, this data product has not been widely used yet.

The present work is a continuation of our previous work \citep[hereinafter, Paper~I,][]{Sharykin2018}. Paper~I was devoted to detailed quantitative multiwavelength analysis of nonthermal electron dynamics and plasma heating in the system of highly sheared magnetic loops interacting with each other above the PIL during the first sub-flare of the confined M1.2 class solar flare on 2015 March 15 (SOL2015-03-15T22:43). This flare was selected because there were emission sources in different ranges of electromagnetic spectrum located in the PIL region with strong photospheric horizontal magnetic field and PVEC. We analyzed X-ray, microwave, ultraviolet and optical observational data. To investigate magnetic structure in the flare region we used a NLFFF extrapolation. To model the microwave gyrosynchrotron emission of nonthermal electrons with the power-law energy distribution (low-energy and high-energy cutoff are 20 keV and 1 MeV, respectively) we used the GX Simulator tool \citep[][]{Nita2015}. It is worth noting that we ignored other two subsequent subflares, as they were characterized by less energetic electrons that were in the focus of the Paper~I. Moreover, the third subflare was mostly thermal flare without nonthermal electrons, or electrons producing HXR emission were under background RHESSI level.

It was found that the observed structure and dynamics of the flare emission sources and magnetic field is consistent with the TCMR scenario. We found that appearance of nonthermal electrons with the hardest spectrum and super-hot plasma ($T \approx 40$~MK) was simultaneous. By simulating the gyrosynchrotron radiation and comparing with observations, it was inferred that accelerated electrons were localized in a thin magnetic channel with a width of around 0.5~Mm elongated along the PIL within the twisted low-lying ($\approx 3$~Mm above the photosphere) magnetic structure with high average magnetic field of about 1200~Gauss. In other words we observed some kind of filamentation of the flare energy release. The accelerated electron density in this energy release region above the PIL was about $10^9$~cm$^{-3}$ (the peak energy rate was about $2 \times 10^{28}$~erg/s) that is much less than the density of the thermal super-hot plasma.
%%%%%%%%%%%%%%%%%%%%%%%%%

Despite the detailed analysis of the multiwavelength observations of the flare emission sources and its modeling done in Paper~I, there was no investigation of dynamics of photospheric magnetic field and PVEC near the PIL. There was also no discussion of electron acceleration and plasma heating as a consequence of magnetic field and PVEC changes around the PIL. It was just mentioned that the observed flare emission sources were located close to the PIL, and the HXR sources were almost co-spatial with the regions of strong PVEC found from the standard (12-minute) HMI vector magnetograms.

The aim of the present work is to perform an extensive quantitative analysis of magnetic field and PVEC dynamics in the SOL2015-03-15T22:43 event using new 135-second HMI/SDO vector magnetograms to study magnetic reconnection near the PIL accompanied by efficient plasma heating and electron acceleration. The main tasks are as follows:
\begin{enumerate}
	\item {to compare positions of the flare X-ray sources and flare ribbons with the photospheric magnetic field components and PVEC;}
	\item {to compare variations of the photospheric magnetic flux and PVEC in the entire PIL region and in the developing flare ribbons with variations of the flare electromagnetic emissions;}
	\item {to estimate the reconnection rate and electric field in the reconnecting current sheet (and some of its physical parameters) to evaluate potential efficiency of the flare energy release;}
\end{enumerate}
Finally, we will discuss the data analysis results in the frame of the TCMR scenario, which agrees well with the observations of this confined flare.

The paper is organized as follows. Section~\ref{S-OBS} describes some general observations of the flare including vector magnetograms, X-ray and optical images. Analysis of HMI 135-second vector magnetograms and NLFFF extrapolations based on these magnetograms are described in Sections~\ref{PVEC135} and \ref{NLFFFsec}, respectively. In Section~\ref{ribbs} we present time profiles of different parameters inside the flare ribbons calculated from 135-second HMI vector magnetograms. The discussion and conclusions of the results obtained are presented in the two final sections.

%%%%%%%%%%%%%%%%%%%%%%%%%%%%%%%%%%%%%%%%%%%%%%%%%%%%%%%%%%%%%%%%%%%%%%%%%%%%%%%%%%%%%%%%%%%%%%%%%%%%%%%%%%%%%%%%%%%%%%%%%%%%%%%%%%%%%%%%%%%%%%%%%%%%%%%%%%%%%%%%%%%%%%
\section{OBSERVATIONS}
\label{S-OBS}
%---------------------------------------------------------------------------------------------------------------------------------------------------------------------
    \subsection{TIME PROFILES OF FLARE EMISSION}

\begin{figure}[!t]
\centering
\includegraphics[width=1.0\linewidth]{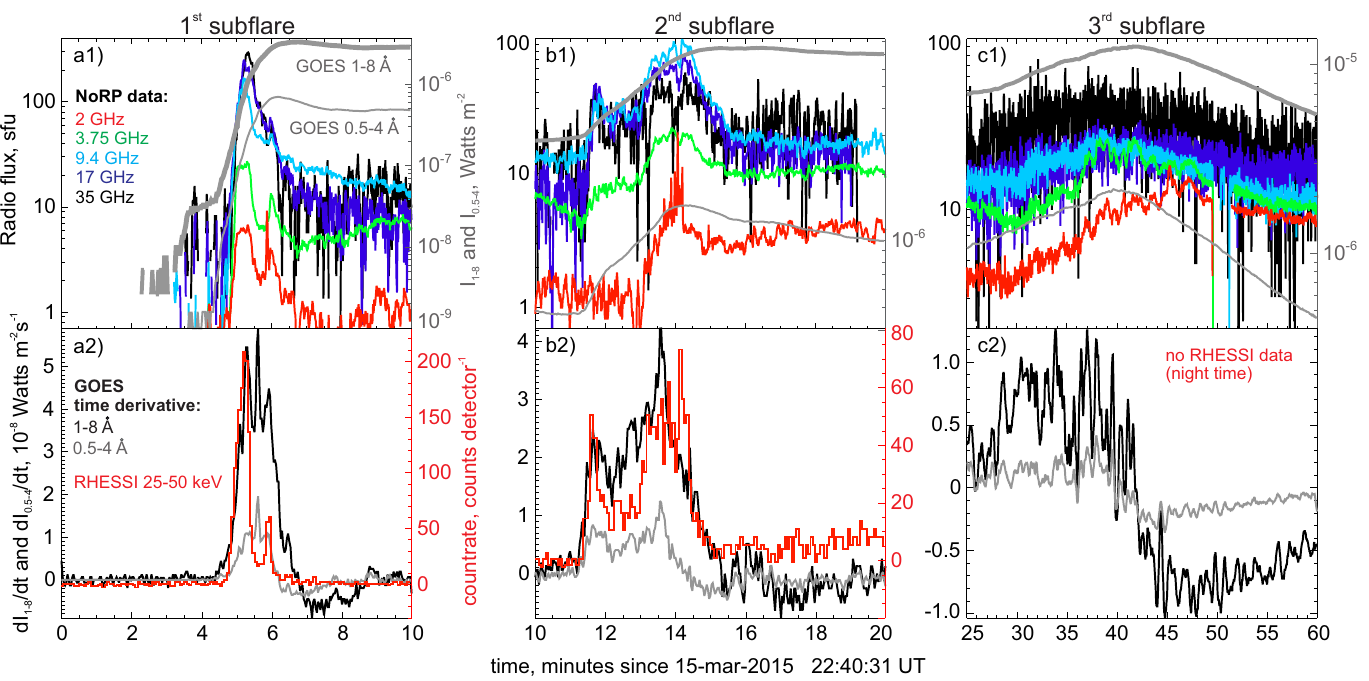}
\caption{Time profiles of the M1.2 solar flare on 15~March 2015 for three subflares (columns a-c). a1-c1) GOES 1-8~\AA{} lightcurve (thick black line; arbitrary units) and NoRP Stokes $I$ time profiles at 2, 3.75, 9.4, 17, and 35~GHz (colors are indicated in the plot). a2-c2) Time derivatives of the GOES lightcurves in both bands (black and grey) and RHESSI lightcurve in the energy band of 25-50~keV in arbitrary units (red histogram). In panel c2 RHESSI data are not plotted due to night time.}
\label{TP}
\end{figure}

Figure~\ref{TP} shows temporal evolution of emission lightcurves of the flare studied. Three columns in figure correspond to three subflares. The top panel demonstrates comparison of the radio fluxes detected by the Nobeyama Radio Polarimeter (NoRP) with the SXR flux in 0.5-4 and 1-8~\AA{} bands detected by the X-ray Sensor onboard the Geostationary Operational Environmental Satellite (GOES). The NoRP time profiles are shown for five frequencies: 2, 3.75, 9.4, 17, and 35~GHz. The bottom panels show the time profiles of the Ramaty High-Energy Solar Spectroscopic Imager (RHESSI; \citep{Lin2002}) 4-second count rate in the energy band of 25--50 keV (red histogram) and of the time derivatives of the GOES X-ray lightcurves. We do not present RHESSI data in c3 panel as the signal was on background level. It can be seen from Figure~\ref{TP} that the three successive sub-flares had approximate duration of the corresponding microwave bursts of about 2--5~minutes. We also should mention that the observed three bursts are considered as parts of the large flare process, as it is shown (see next sections) that all emission sources and changes of the magnetic field was in the same region near the PIL. Moreover, UV maps \citep[][]{Sharykin2018} revealed continual transition of the energy release from the decay phase of the previous sub-flare to the impulsive phase of the next sub-flare. In other words, the subsequent sub-flares are developed from the initial magnetic configurations prepared by the previous sub-flare energy release.

The detected HXR and microwave emissions were generated by accelerated electrons. Intensities of these emissions were maximal during the first sub-flare, which was investigated in details in Paper~I. The spectral analysis (performed in Paper I) of the X-ray emission detected in the first sub-flare has revealed that the spectrum consists of the two main components: thermal ($\lesssim 20$~keV) and nonthermal ($\gtrsim 20$~keV). By thermal and nonthermal we mean the spectral components, which can be fitted by the Maxwellian and power-law distributions, respectively. In particular, the detailed modeling of the microwave gyrosynchrotron emission was performed in Paper I, and we remind that the dominating emission at 35~GHz during the first sub-flare is a result of the gyrosynchrotron radiation of nonthermal electrons localized in the low-lying magnetic loops stretched along the PIL with the average magnetic field $\approx 1200$~G. Figure~\ref{TP} also reveals that the HXR and microwave emission peaks correspond to peaks in the time derivative of the GOES lightcurves \citep[known as the Neupert effect; ][]{Neupert1968}. Further, in this paper, we will use the time derivative of the GOES lightcurve in 1-8~\AA{} band to compare qualitatively the time evolution of the flare energy release process with the dynamics of the photospheric magnetic fields and PVEC.

To sum up, the selected flare reveals non-stationary energy release process with the three stages characterized by different efficiency of plasma heating and acceleration of electrons. Paper~I was concentrated on the study of the first sub-flare, as it reveals the most intense HXR and microwave emission sources in the vicinity of the PIL that allowed to investigate accelerated electrons and plasma heating at the PIL in details. However, in this work we will investigate dynamics of the magnetic field and PVEC for the entire flare including all three sub-flares. We know from the HXR emissions that the first subflare was more impulsive and energetic (HXR emission with the hardest spectrum) than the second subflare which was more energetic than the third subflare. Further we will compare magnetic field dynamics in the flare energy release sites with the onsets of the subflares and try to answer the question what determines the energy release efficiency (acceleration and heating rate).

%---------------------------------------------------------------------------------------------------------------------------------------------------------------------
   \subsection{COMPARISON OF X-RAY EMISSION SOURCES WITH MAGNETIC FIELD AND PVEC}

\begin{figure}[!t]
\centering
\includegraphics[width=1.0\linewidth]{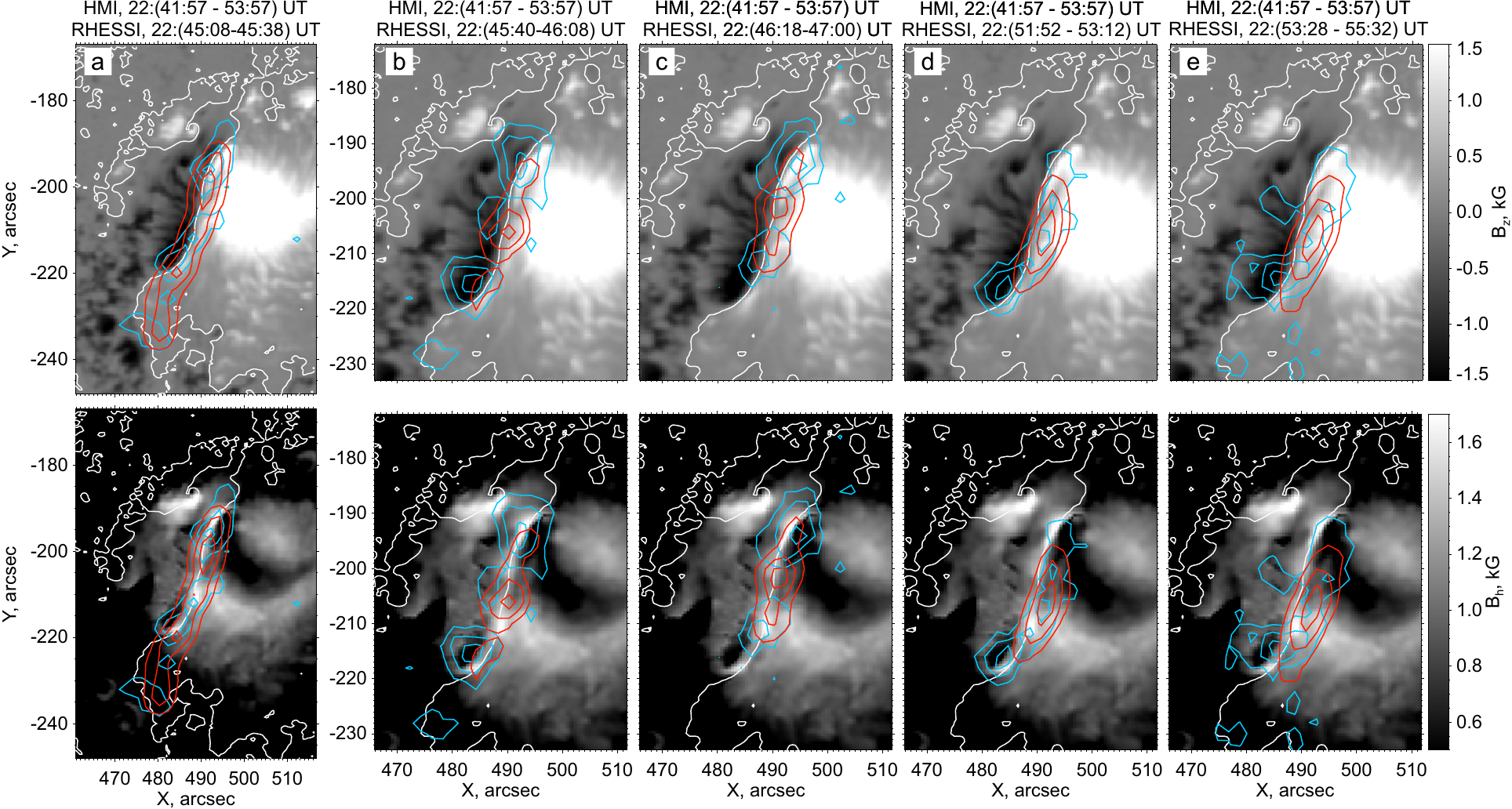}
\caption{Location of the X-ray sources detected by RHESSI in the $6-12$ (red) and $25-50$~keV (light blue) energy bands in three time intervals (from left to right) during the first subflare of the 15 March 2015 flare studied. Contours correspond to 50, 70, and 90 \% levels of the maximal X-ray intensity for each map. The corresponding time intervals are indicated above the top panels. Top and bottom panels present HMI vector magnetograms showing distributions of the vertical and horizontal magnetic field components on the photosphere, respectively. The photospheric PIL is shown by the white curves. Note that the field-of-view size is centered and adjusted relative to the RHESSI contours and slightly differs for the left and other four panels.}
\label{RHESSI_B}
\end{figure}

\begin{figure}[!t]
\centering
\includegraphics[width=1.0\linewidth]{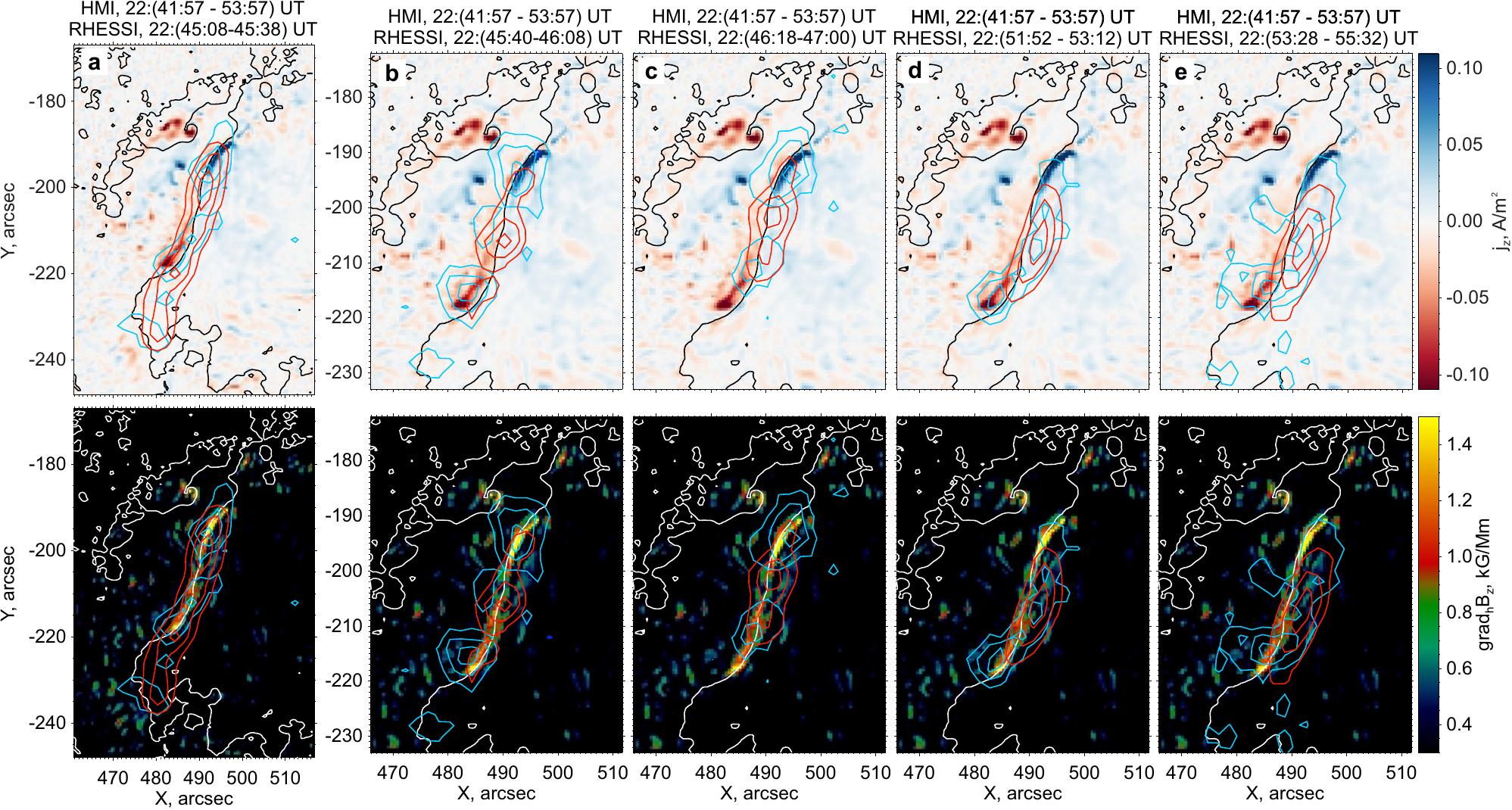}
\caption{Comparison between the RHESSI X-ray contour maps in the energy bands of 6-12 (red) and 25-50~keV (light blue) for three different time intervals (indicated above the top panels) and the maps showing distribution of the vertical electric currents (top panels) and horizontal gradient of vertical magnetic field component (bottom panels) deduced from the 720-second HMI vector magnetogram. The photospheric PIL is shown by the black (top) and white (bottom) curves. Contours correspond to 50, 70, and 90 \% levels of the maximal X-ray intensity for each map. Note that the field-of-view size is centered and adjusted relative to the RHESSI contours and slightly differs for the left and other four panels.}
\label{RHESSI_j}
\end{figure}

The X-ray maps (Figures~\ref{RHESSI_B} and \ref{RHESSI_j}) were reconstructed with the CLEAN algorithm \citep{Hurford2002} using the data of the RHESSI detectors 1, 3, 5, 7, and 8 for three time intervals during the first sub-flare (panels a-c) and two time intervals for the second sub-flare (panels d, e). We do not present maps for the third sub-flare due to the low RHESSI count rate above 25~keV that leads to too noisy images. Figure\,\ref{RHESSI_B} demonstrates the X-ray sources in two energy bands of 6--12 and 25--50~keV shown by the contours at 50, 70 and 90 \% of the maximum of each map. The first band corresponds mostly to the thermal emission, while the second one is mostly produced by accelerated (nothermal) electrons precipitated into the chromosphere. The X-ray maps are plotted as the contours of constant levels and overlayed onto the HMI magnetic field maps. For this comparison we used the standard HMI vector magnetogram with temporal resolution of 720~s, as they are less noisy than 135-s ones. Vector magnetic field components were calculated from the reprojected onto the heliographic grid HMI vector magnetogram in the cartesian heliocentric coordinates. Vertical and horizontal magnetic field maps are shown in the top and bottom panels of Figure\,\ref{RHESSI_B}, respectively. The PIL has the S-shaped form that is typical for active regions with strong electric currents and containing sigmoids.

As visible in Figure~\ref{RHESSI_B}, thermal and nonthermal HXR emission sources were generated close to the PIL (see top panels), where the horizontal magnetic field dominates (see bottom panels) and reaches values of 1500~Gs. In the beginning of the impulsive phase (panel a) there were three major non-thermal HXR sources with different intensities, and the thermal X-ray source at this time had an elongated (along the PIL) shape. The thermal X-ray emission was generated from the thin channel with the approximate length of $\approx 40^{\prime\prime}$, filled with hot plasma. From the Paper~I we also know that this elongated X-ray emission source is also seen in the ``hot'' 94 and 131~\AA{} AIA channels (which are sensitive to emission of flaring plasmas with temperatures $\log_{10}{T} \approx 6.8-7.0$) and is associated with the strongly sheared magnetic loops. Thus, the observed EUV and X-ray emissions were generated, most probably, from the same flare loops. The X-ray contour maps in the subsequent two time intervals (panels b, c) show the rather compact thermal X-ray source located between the two strongest non-thermal HXR sources. In these cases we observe the highly-sheared (about 80 degree) magnetic structure. During the second sub-flare (panels d, e), we observed two major non-thermal HXR sources on opposite sides of the PIL. However, the single elongated thermal source was not located exactly at the PIL, as in the case of the first sub-flare. It was slightly ($\approx 2^{\prime\prime}$) shifted west. This could be a result of the two effects: (1) the expansion of the flare energy release region leading to the apparently upward displacement of the loop-top thermal X-ray source, and (2) the projection effect (since the flare was not exactly in the center of the solar disk).

To reconstruct vertical PVEC density ($j_{z}$) maps shown in the top panels of Figure\,\ref{RHESSI_j} we applied Ampere's law to the horizontal magnetic field components in the HMI vector magnetograms \citep[e.g.][]{Guo2013,Musset2015}. The two ribbon-like regions of the highest PVEC density are located near the PIL and close to the two strongest non-thermal HXR sources (the intense PVEC regions are covered by 50\% HXR contours) for both considered sub-flares. However, similar to the previous works, there are no exact spatial correspondence, i.e. the most intense HXR pixels are not co-spatial with the pixels of the highest PVEC density, at least for some time intervals considered. The best correspondence between the brightest HXR pixel and the nearest HMI pixel with the locally maximal PVEC density was at the initial time (panel~a). The southern strongest HXR source almost coincided with the strongest PVEC region (the displacement was less than 1$^{\prime\prime}$). But the displacement of the northern strongest HXR source from the HMI pixel with the locally maximal PVEC value was about 3$^{\prime\prime}$. For the panel b, the displacement distance is about 3$^{\prime\prime}$ for the southern HXR source and 5$^{\prime\prime}$ for the northern one. The largest displacement of the HXR source from the region of the locally maximal PVEC density was at the time of panel~c for the southern source, when the distance was about 7$^{\prime\prime}$ and strongest PVEC region was out from the 50\% HXR contour. At later times, it is difficult to estimate the distances between the local maxima of PVEC density and non-thermal HXR sources due to their relatively low brightness and increased noise in the synthesized images.

In the bottom panels of Figure\,\ref{RHESSI_j} we show the absolute value of the vertical magnetic field gradient $\nabla_{h}B_{z} = \sqrt{(\partial B_{z}/\partial x)^2+(\partial B_{z}/\partial y)^2}$ in the local plane of the solar surface. The strongest values of $\nabla_{h}B_{z}$ ($\approx 1-2$~kG/Mm) are found along the PIL, and also concentrated in the vicinity of the two strongest non-thermal HXR sources. The best coincidence between the HXR sources and the regions of the strongest $\nabla_{h}B_{z}$ was achieved in the beginning of the flare impulsive phase (panel~a). The best coincidence between the non-thermal HXR sources (brightest pixels) and maximal $\nabla_{h}B_{z}$ was for the southern sources in panels~a,d and for the northern sources in panels a,b (with displacements $\lesssim 1^{\prime\prime}$). In other cases, the displacement values varied in the approximate range of $2^{\prime\prime} - 7^{\prime\prime}$ that is similar for the PVEC maps.

\begin{figure}[!t]
\centering
\includegraphics[width=1.0\linewidth]{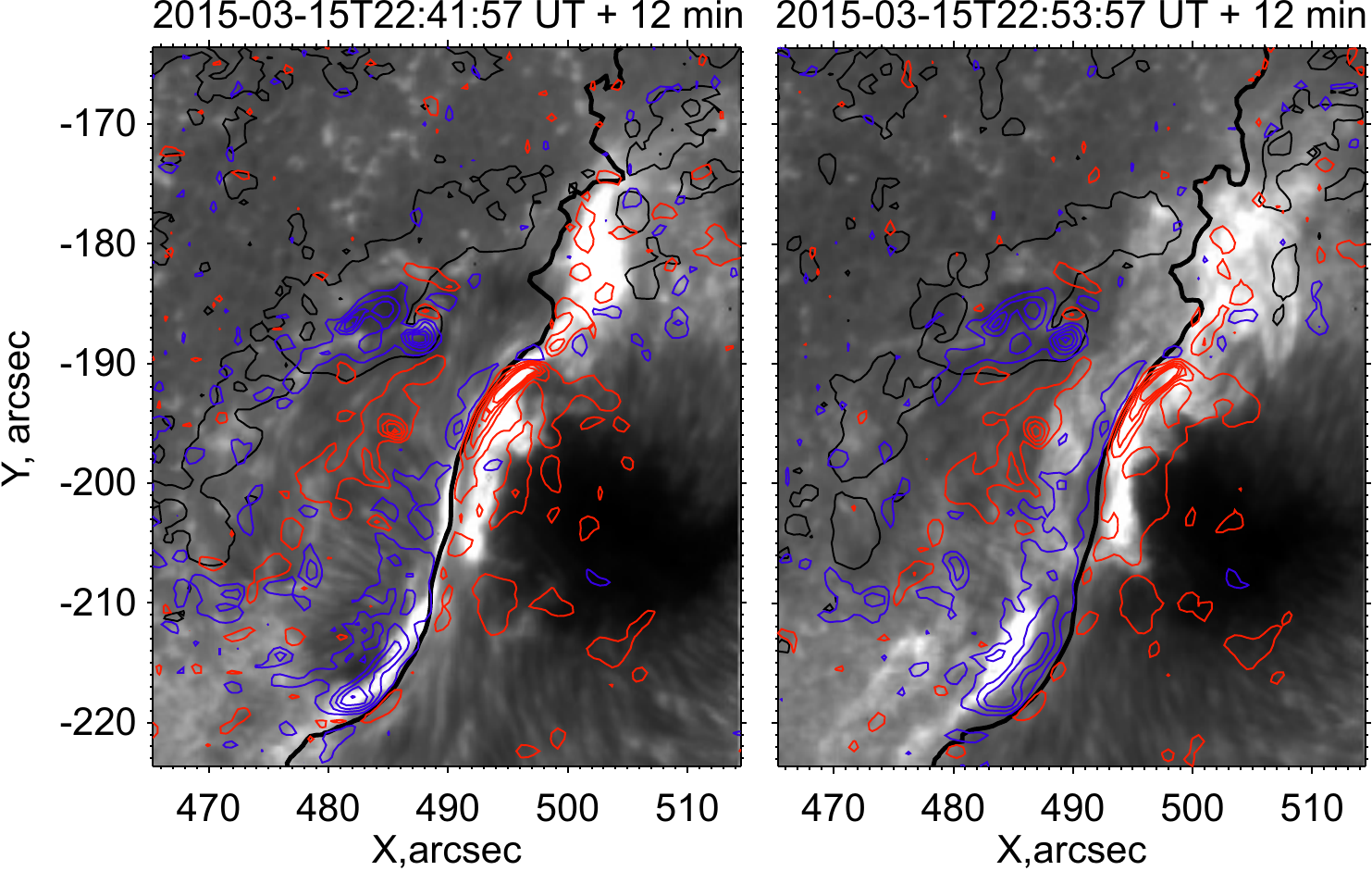}
\caption{Comparison between Ca II SOT/Hinode cumulative images (background) and contour maps of the PVEC and the PIL (black curve) deduced from the HMI vector magnetograms. The left and right panels show two subsequent time intervals. The red and blue contours correspond to the negative and positive PVEC density (15, 45, 75, 105, and 135~mA/m$^2$), respectively. The SOT cumulative image is a result of summing Ca II images taken in time interval of 12 min corresponding to the HMI vector magnetograms.}
\label{SOT_j}
\end{figure}

To resolve fine spatial structure of the flare energy release in the lower solar atmosphere and to compare it with the PVEC density maps we used Ca~II ($\lambda=$3968.5~\AA{}, this emission is formed in the lower chromosphere) images from the Solar Optical Telescope (SOT: \cite{Tsuneta2008}) onboard the space solar observatory Hinode \citep{Kosugi2007}. In Figure\,\ref{SOT_j} we present comparison between two $j_{z}$ maps, deduced from two successive 720-second HMI vector magnetograms, and two optical cumulative images from SOT. The cumulative image is a sum of all available SOT images within the corresponding 720-second integration time interval of the corresponding HMI magnetogram. Such images present information about spatial distribution of total photospheric response during the flare impulsive phase and allow to compare roughly the high-cadence Hinode data with the HMI maps with the low temporal resolution. The most intense optical emission was generated in the regions of enhanced PVEC density, and the most intense emission sources had the best coincidence with the strongest PVEC during the first 12-min time interval (left panel), but were out of the places with the highest $j_{z}$ values during the subsequent 12-min interval (right panel).

To sum up, in general, the flare non-thermal HXR and optical emission sources were in the vicinity ($\approx 2^{\prime\prime} - 7^{\prime\prime}$) to the regions of enhanced PVEC and at some time intervals, especially in the begin of the flare impulsive phase, the localized regions of the most intense emissions were very close (within $\approx 1^{\prime\prime}$) to the strongest PVEC.

%%%%%%%%%%%%%%%%%%%%%%%%%%%%%%%%%%%%%%%%%%%%%%%%%%%%%%%%%%%%%%%%%%%%%%%%%%%%%%%%%%%%%%%%%%%%%%%%%%%%%%%%%%%%%%%%%%%%%%%%%%%%%%%%%%%%%%%%%%%%%%%%%%%%%%%%%%%%%%%%%%%%%%
\section{CHANGES OF PHOTOSPHERIC MAGNETIC FIELD AND PVEC AROUND THE PIL FROM HMI VECTOR 135-SECOND MAGNETOGRAMS}
\label{PVEC135}

In the previous section we compared the X-ray and optical images with the magnetic field maps deduced from the HMI vector magnetograms with the time cadence of 720~s. This temporal resolution is insufficient to resolve magnetic field dynamics during the sub-flares, whose impulsive phase has typical duration of $\approx 5$ minutes. In this section, we will describe the magnetic field dynamics during the flare studied using the high-cadence HMI vector magnetograms with the temporal resolution of 135~s.

The HMI 135-seconds vector magnetograms were previously described by \cite{Sun2017}. It was shown that the maximum of magnetic field distribution is achieved at $\approx 130$~Gs. It is typical for the quiet Sun, as the polarization degree is low, and the most probable value is due to the noise. This noise mostly affects the determination of the horizontal magnetic field component. We reconstructed distribution of magnetic fields in the quiet Sun outside the active region where the flare was triggered and found the most probable value is 110~Gs. This value was taken as the upper boundary for the magnetic strength error. Due to the summation, the magnetic flux error is very small for the flare regions composed of many pixels and negligible compared to the flux values.

%Let's consider magnetic flux from the region consisting of $N$ HMI pixels. In this case the magnetic flux error will be proportional to $\sqrt{N}$ and thus is very small for the large-scale flare regions (since the magnetic flux is proportional to $N$) even for the horizontal magnetic field component. We also used the Monte-Carlo simulation to estimate errors of calculated magnetic fluxes and found them very small. Further in the work we will not show the error bars for magnetic fluxes as they are small comparing with the size of the data points. Variations in the time-profiles for the magnetic field components and PVEC are mostly due to the limited region-of-interest (ROI) considered. We do not track the isolated magnetic flux tube. Different types of horizontal motions can lead to irregular variations for calculated fluxes inside the small ROI considered.

We should also mention that the observed dynamics of the magnetic fields and PVEC during the flare studied is not distorted by artifacts connected with the wrong measurements of the Fe~I line. Such incorrect HMI data could be connected with sharp energy release in the lower solar atmosphere leading to distortion of the Stokes profiles of the measured line. The argument that it is not our case is the absence of fast and high amplitude variations from pixel to pixel on the magnetograms analyzed. Such variations are quite typical for strong white-light solar flares. For example, in the work of \cite{Sun2017} all 4 Stokes profiles were compared for the flare ribbon and quite Sun regions. There were strong distortions in the flare ribbon that explains magnetic field artifacts. We also checked the Stokes profiles for the selected points in the vicinity of the PIL, where HXR, UV and optical emissions were generated. We found that the Fe~I line did not reveal strong distortions, and, thus, we assume that the magnetic field is measured correctly. Therefore, PVEC is also calculated correctly. We also emphasize here that the flare studied was quite a weak M1.2 class flare.

\begin{figure}[!t]
\centering
\includegraphics[width=1.0\linewidth]{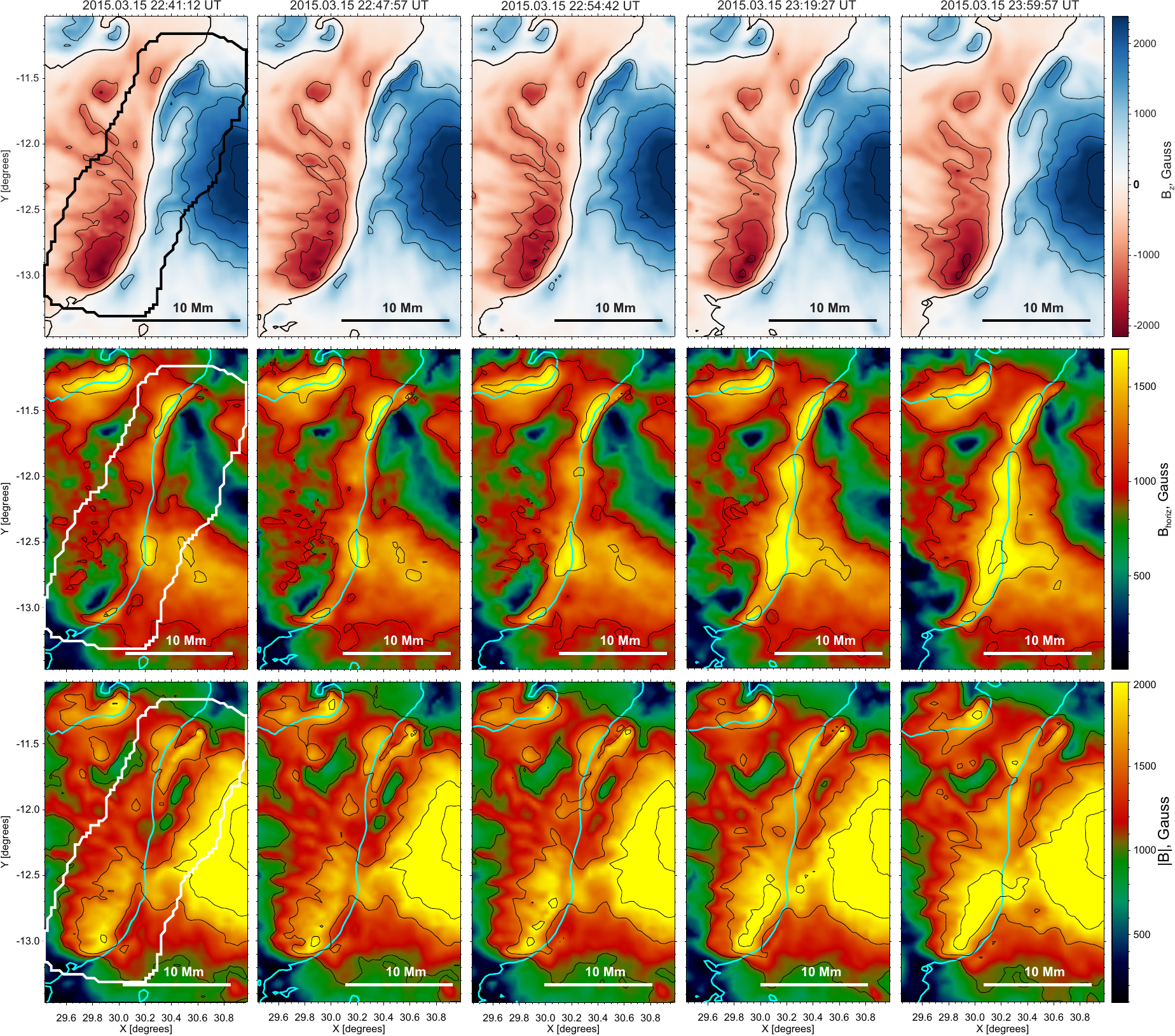}
\caption{Time sequence (time is growing from left to right) of magnetic field component maps, deduced from HMI 135-second vector magnetograms and reprojected onto the heliographic grid. The time is indicated above the top panels. The top, middle and bottom panels correspond to the vertical, horizontal magnetic component and absolute value of the magnetic field, respectively. The photospheric PIL is shown by the black (top) and cyan (middle, bottom) curves. The black and white contours in the left column show the region-of-interest (similar to the black contour in Figure~\ref{Idyn}(a) where we calculated magnetic fluxes shown in Figure~\ref{Bdyn}.}
\label{Bmaps}
\end{figure}

\begin{figure}[!t]
\centering
\includegraphics[width=1.0\linewidth]{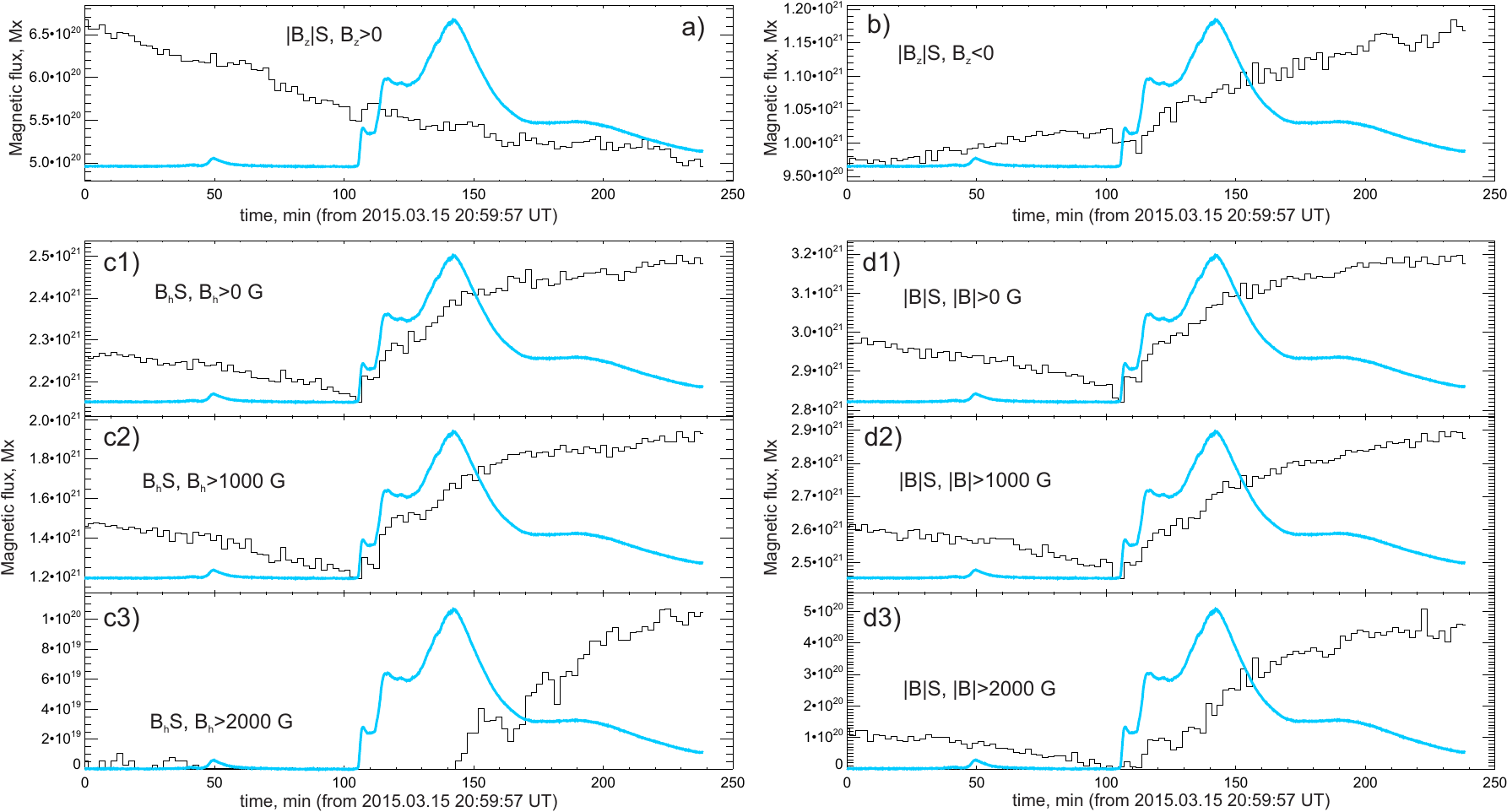}
\caption{Time profiles of magnetic fluxes in the PIL region around the strongest PVEC. To compare with the real magnetic flux $B_zS$, we introduced the nominal magnetic flux for the horizontal component and for the absolute value of the magnetic field. The analyzed region is shown by the black contour in Fig.\,\ref{Idyn}(a). Panels show the following information: (a) and (b) magnetic fluxes for the vertical magnetic field components with positive and negative sign, respectively. (c1--c2) Nominal magnetic fluxes for the horizontal component summing pixels with $B_{h}>0$, 1000, and 2000~G, respectively. (d1--d2) Nominal magnetic fluxes the absolute value summing pixels with $|B|>0$, 1000, and 2000~G, respectively. The blue curve in panels (a--d) is the GOES X-ray lightcurve (1-8~\AA) in arbitrary units.}
\label{Bdyn}
\end{figure}

\begin{figure}[t]
\centering
\includegraphics[width=0.67\linewidth]{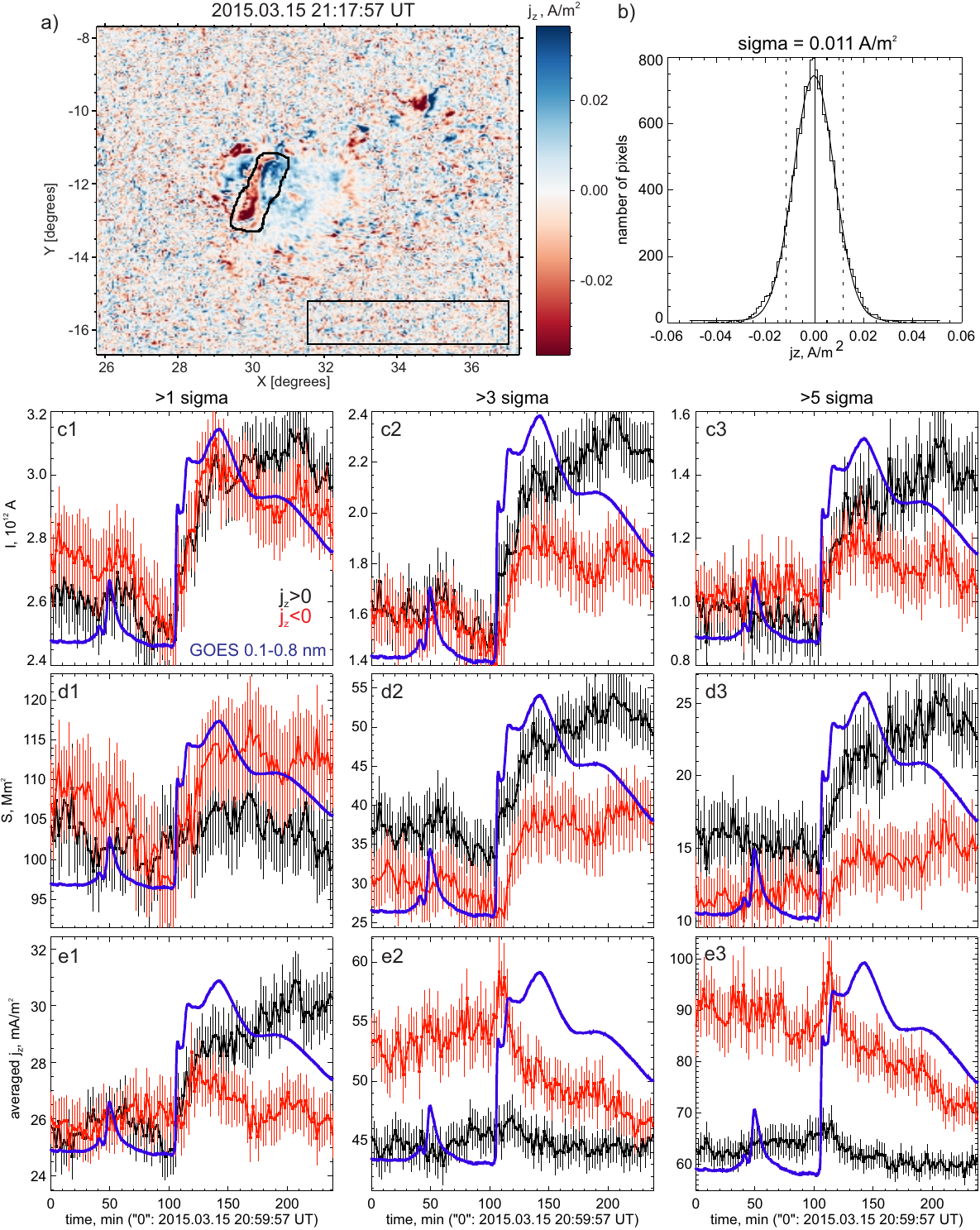}
\caption{Temporal dynamics of the total PVEC $I_{z}$ in the flare region (marked by the thick black contour in the panel (a) is shown in the panels (c1)--(c3). Temporal dynamics of the total area ($S$) of the regions with the enhanced PVEC is shown in the panels (d1)--(d3). The panels (e1)--(e3) present temporal profiles of the estimated averaged (or effective) PVEC density (the ratio of the total PVEC $I_{z}$ to the total area $S$). The measurement errors are shown by the vertical bars. Three panels (from left to right) in each raw (c--e) show cases considering the PVEC density above the threshold values $1\sigma(j_{z})$, $3\sigma(j_{z})$, and $5\sigma(j_{z})$, respectively. The thin black rectangle in the right-bottom corner of the panel (a) corresponds to the non-flaring (background) region where we calculate the PVEC distribution to determine the noise level $\sigma(j_{z})$. This distribution is shown in the panel (b) by the histogram, where the solid line is a Gaussian fit. The blue curve in panels (c--e) is the GOES X-ray lightcurve (1-8~\AA) in arbitrary units.}
\label{Idyn}
\end{figure}

The time sequence of the magnetic field component maps is shown in Figure\,\ref{Bmaps}. There are maps for the magnetic field absolute value (bottom panels), horizontal (middle panels) and vertical (top panels) components. These maps reveal that there were no significant changes in the vertical magnetic field component, while the horizontal magnetic field near the PIL was significantly intensified during the flare. Figures\,\ref{Bdyn}(a,\,b) demonstrate the time profiles of magnetic fluxes $F_z = |\sum B^{ij}_{z}|S_{pix}$ for the negative and positive vertical components, where $ij$ superscript means summation through all pixels inside a particular region and $S_{pix}$ is the single pixel area. These fluxes were calculated in the area (shown by the black thick contour in Figure\,\ref{Idyn}(a)) around the PIL with the high PVEC density. Both positive and negative magnetic fluxes do not reveal significant changes around the flare onset. The negative and positive fluxes have opposite trends: decreasing and increasing, respectively.

To compare real magnetic flux $F_z$ (integrated over the region-of-interest, ROI, for $\vec{B}\cdot \vec{n}$, where $\vec{n}$ is the normal vector to the photosphere) with dynamics of the other magnetic field components through all analyzed area, we introduce the nominal magnetic fluxes for the horizontal component ($B_h$) and for the absolute value of the magnetic field ($|B|$), same as for the vertical magnetic component. Figures\,\ref{Bdyn}(c1--c3) show temporal dynamics of magnetic flux $\sum B^{ij}_{h}S_{pix}$ for the horizontal magnetic field component. These three panels present the time profiles of the fluxes calculated by summing pixels with magnetic field values higher than three thresholds: of 0, 1, and 2~kG. The relative magnetic flux change in Figure\,\ref{Bdyn}(c1) $(F^h_{max}-F^h_{min})/F^h_{min} \approx 0.4 \times 10^{21}/2.1\times 10^{21}\approx 0.19$, where $F^h_{min}$ and $F^h_{max}$ are the magnetic fluxes before the flare onset (the minimal value) and after the flare (the maximal value), respectively. The relative flux change in Figure\,\ref{Bdyn}(c2) is about 0.58 with $F^h_{max}\approx 1.9\times 10^{21}$~Mx which is 76\,\% from the total horizontal magnetic flux. In other words, the largest fraction of the horizontal magnetic flux is contained in the strong magnetic field with $B_{h}>1000$~G. The strongest ($B_{h}>2000$~G) horizontal magnetic field appeared around the major peak (the third sub-flare) of the flare X-ray flux, approximately 40~minutes after the flare onset (Figure\,\ref{Bdyn}(c3)).

The time profiles for the nominal total magnetic fluxes $\sum |B^{ij}|S_{pix}$ ($|B| = \sqrt{B_{h}^2+B_{z}^2}$) are plotted in Figures\,\ref{Bdyn}(d1--d3). The fluxes shown in Figures\,\ref{Bdyn}(d1--d3) were calculated, as before, by summing pixels with the magnetic field absolute values higher than three thresholds: of 0, 1, and 2~kG. The change of the total magnetic flux is also associated with the flare onset. According to Figure\,\ref{Bdyn}(d1), the maximal total magnetic flux $F_{max} \approx 3.2 \times 10^{21}$~Mx and the relative change is $\approx 0.12$. Comparing panels (c) and (d) we can conclude that the largest fraction of the total magnetic flux is contained in the horizontal magnetic field. For example, comparing panels (c1) and (d1) we have $F^h_{max}/F_{max}\approx 0.78$, or comparing panels (c2) and (d2) $F^h_{max}/F_{max}\approx 0.59$. We also found that more than half (53\,\%) of the total magnetic flux (d1) is in the strong ($>1000$~G) horizontal magnetic field (c2).

\begin{figure}[t]
\centering
\includegraphics[width=1.0\linewidth]{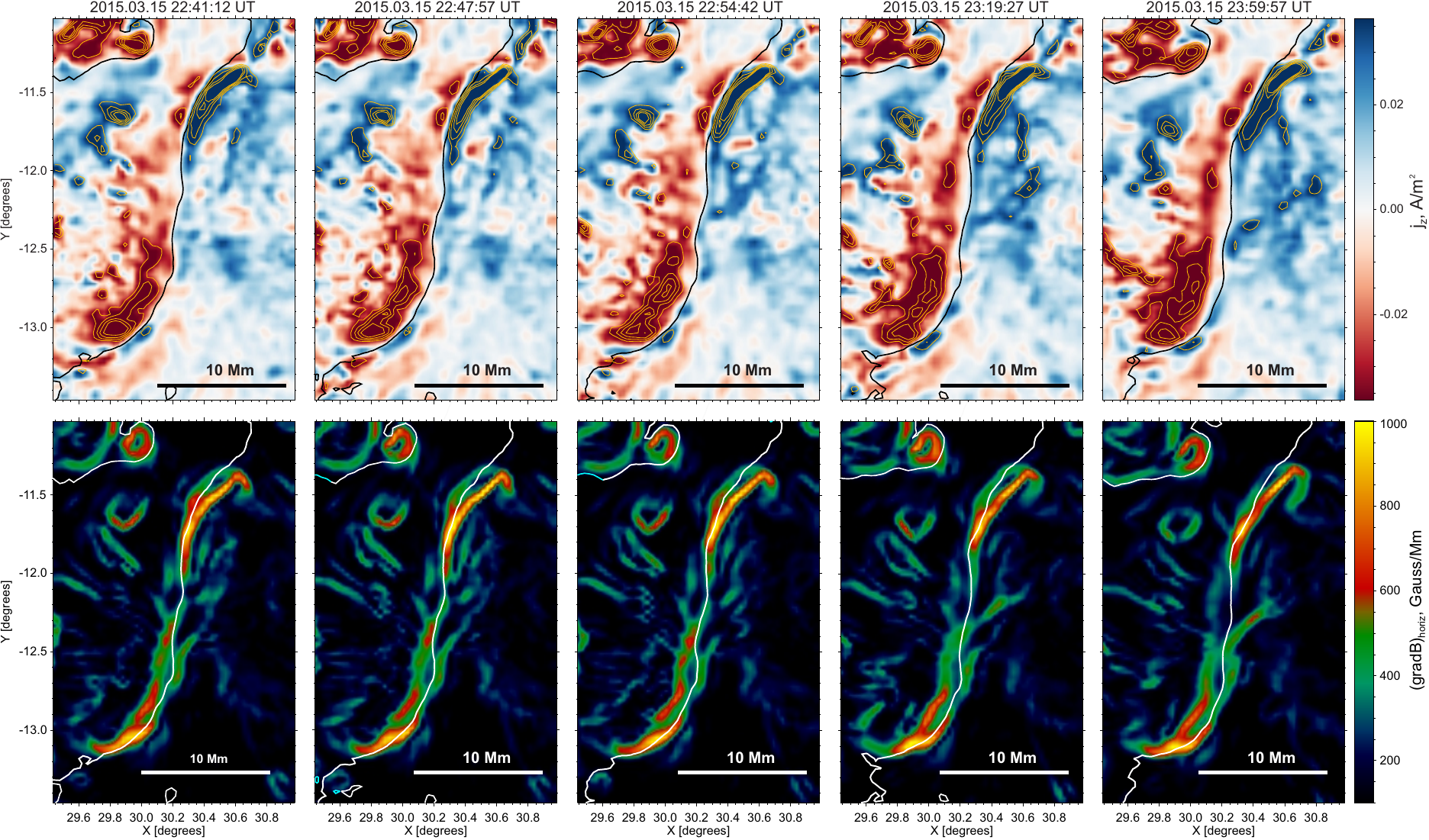}
\caption{Temporal sequence (corresponding time is indicated above the top pannels) of maps showing spatial distributions of PVEC (top panels) and horizontal gradient of the vertical magnetic field component (bottom panels). These maps are deduced from the HMI 135-sec vector magnetograms and reprojected onto the heliographic grid. The photospheric PIL is shown by the black (top) and white (bottom) curves. The orange contours mark PVEC density levels of 39, 62, 85, and 115 mA/m$^2$ for both $j_z$ signs.}
\label{jmaps}
\end{figure}

The spatial distributions of PVEC density and horizontal gradient of vertical magnetic field around the PIL are shown in Figure\,\ref{jmaps} for five time instants. Visual analysis of these figures does not show obvious changes. To demonstrate changes in the PVEC system quantitatively we calculated the average characteristics (see Figure\,\ref{Idyn}) for the pixels with the enhanced PVEC density in the total PIL region, marked by the black thick contour in panel (a). Three panels (from 1 to 3) in each raw (c--e) show cases considering the pixels with PVEC density above three thresholds with values of $1\sigma(j_{z})$, $3\sigma(j_{z})$ and $5\sigma(j_{z})$. The black and red colors in each panel of Figure\,\ref{Idyn}(c--e) correspond to the positive and negative PVEC, respectively.

%The top panels reveal variations in the spatial structure of the PVEC density. One can notice that the region with strong PVEC density experienced slight expansion around the PIL. It can be seen for the outer contours (especially for the negative PVEC shown by red). Another peculiarity is fragmentation of the region with strong PVEC density. In the first map there was only one centroid for the negative PVEC density in the south part of the PIL. The later times reveal three strong intensification. In the case of the magnetic gradient map there were no significant changes. The regions with the largest gradients experienced slight distortions.

To estimate the background (noise) level $\sigma(j_{z})$ for the PVEC density we calculated the $j_{z}$ distribution in the non-flaring region marked by the black rectangle in the lower right corner in Figure\,\ref{Idyn}(a). This distribution is shown in Figure\,\ref{Idyn}(b) by the histogram, where the solid line is a Gaussian fit and the two dashed vertical lines mark the $1\sigma(j_{z})$ level. The calculated sigma is about 11~mA/m$^2$, which is comparable with the estimation $\sigma_j = c\sqrt{6}\sigma_B/(16\pi\Delta x)\approx 14\pm4$~mA/m$^2$ for $\sigma_B\approx 110\pm30$~G and pixel size $\Delta x = 0.5^{\prime\prime}$. Here $c$ is the speed of light in vacuum. To estimate errors in determination of the total PVEC (c1--c3) in the ROI, the area of the ROI (d1--d3) with the enhanced PVEC density, and the effective PVEC density (or, by the other words, the PVEC density averaged over the ROI; e1--e3) we used the Monte-Carlo simulation technique. In the selected ROIs we added Gaussian noise with $\sigma_j = 11$~mA/m$^2$ to calculate $I_z$ map and deduced all needed parameters in 100 runs. Then, by calculation of the standard deviations, we found all needed sigmas for each time interval, and show the errors in Figure~\ref{Idyn}(c1-e3) by the vertical bars of three sigmas length.

%The small amplitude variations are in the range of the three sigma level. However, some monotonic variations on the time scale of a few 135-s frames can be due to the limited ROI and connected with the photospheric motions.

The temporal dynamics of the total PVEC $I_{z}$ in the PIL region is shown in Figure\,\ref{Idyn}(c1--c3). $I_{z}$ was calculated as the sum of PVEC density values for pixels inside the ROI and multiplied by the pixel area. One can see that $I_{z}$ had a jump during the flare onset (the first sub-flare) and continued increasing --- until around the major flare SXR emission peak for $I_{z}<0$ and even further for $I_{z}>0$. Considering the case $|j_{z}|>1\sigma(j_{z})$ shown in Figure\,\ref{Idyn}(c1), the relative value of $I_{z}$ change is estimated as $[max(I_{z})-min(I_{z})]/min(I_{z}) \approx 0.26$ for both polarities. The higher amplitudes were achieved in the cases $|j_{z}|>3\sigma(j_{z})$ in Figure\,\ref{Idyn}(c2) and $|j_{z}|>5\sigma(j_{z})$ in Figure\,\ref{Idyn}(c3): 0.41 and 0.22 (for $j_{z}<0$), 0.69 and 0.73 (for $j_{z}>0$), respectively.

Temporal dynamics of the total area of the regions with the enhanced PVEC density is shown in Figure\,\ref{Idyn}(d). The area was calculated as a sum of all pixels area above the corresponding thresholds of $1\sigma(j_{z})$, $3\sigma(j_{z})$, and $5\sigma(j_{z})$. The changes of the total area similar to $I_{z}$, i.e. the increasing after the flare onset. Figures\,\ref{Idyn}(d2) and (d3) reveal the largest jumps of the area for the positive PVEC density: from 26 to 48 and from 12 to 20 Mm$^2$, respectively. Thus, the total PVEC in the ROI increased simultaneously with increasing photospheric cross-sectional area of the electric current carrying magnetic structure.

Figures\,\ref{Idyn}(e) present the temporal profiles of the estimated average (or effective) PVEC density $<j_{z}>$ in the PIL region. It is calculated as the ratio of the total PVEC $I_{z}$ (Figures\,\ref{Idyn}(c)) to the area (Figures\,\ref{Idyn}(d)). The changes of $<j_{z}>$ have different trend comparing with the total PVEC and area of the regions with enhanced PVEC density. There is a peak of $<j_{z}>$ after the flare onset. This increase of $<j_{z}>$ is especially pronounced for $j_{z}<0$ shown in Figure\,\ref{Idyn}(e2) and Figure\,\ref{Idyn}(e3), where $|max(<j_{z}>)| \approx 70$ and $\approx 105$~mA/m$^{2}$, respectively. After this peak, $<j_{z}>$ gradually decreased with some weak fluctuations. One can see this tendency in each panel of Figures\,\ref{Idyn}(e) for both current signs, except for $|j_{z}|>1\sigma(j_{z})$ (for $j_{z}>0$) in Figure\,\ref{Idyn}(e1) that could be due to contribution of the background into $<j_{z}>$ estimation.

To sum up this section, the high-cadence 135-second HMI magnetograms helped to reveal the intensification of the horizontal magnetic field component around the PIL, where we observed the flare emission sources. The vertical magnetic field component did not show flare-related dynamics. The total PVEC $I_{z}$ in the flare region around the PIL gradually increased during the entire flare, while the estimated averaged PVEC density showed non-monotonic dynamics with the peak during the first sub-flare and decreasing trend after that.

%%%%%%%%%%%%%%%%%%%%%%%%%%%%%%%%%%%%%%%%%%%%%%%%%%%%%%%%%%%%%%%%%%%%%%%%%%%%%%%%%%%%%%%%%%%%%%%%%%%%%%%%%%%%%%%%%%%%%
\section{3D MAGNETIC STRUCTURE OF THE FLARE REGION}
\label{NLFFFsec}

The HMI vector magnetograms allow to investigate magnetic field dynamics only on the photospheric level. To study temporal dynamics of the 3D coronal magnetic field structure we used the NLFFF extrapolation applied to the time sequence of the HMI 135-second vector magnetograms used as the boundary condition. The magnetic field extrapolation was made using the implementation of the optimization algorithm \citep{Wheatland2000} developed by \citet{Rudenko2009}. The same procedure and extrapolation parameters are used as in Paper~I (see Paper~I for details). It should be noted here that the flare studied was a confined event without eruption of a filament and magnetic flux-rope. Consequently, the flare was not accompanied by destruction of the magnetic structure of the active region. This partly justifies the use of a force-free approach to describe the magnetic structure and its dynamics in a given event.

\begin{figure}[t]
\centering
\includegraphics[width=1.0\linewidth]{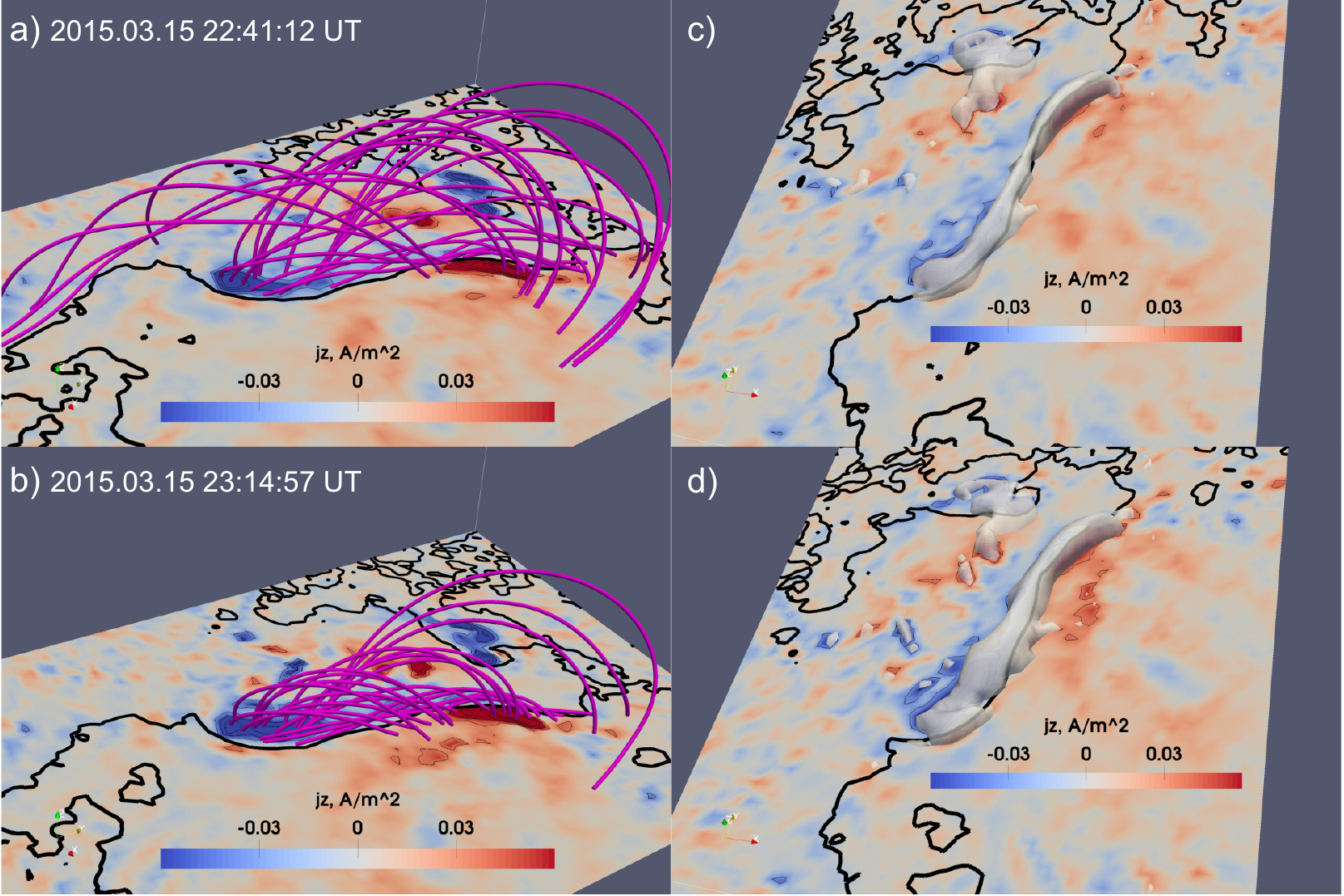}
\caption{Magnetic field lines in the PIL region calculated from the NLFFF magnetic field extrapolations using the HMI 135-sec vector magnetograms as the boundary condition for two different times just before the flare impulsive phase (a) and during main flare X-ray emission peak (b). The regions of strong electric currents flowing along the magnetic field lines for the same times are shown by the white semi-transparent surfaces with the constant level of electric current density of 55~mA/m$^{2}$ in (c) and (d), respectively. The base maps show the distributions of vertical electric current density on the photosphere by blue-red (negative-positive) color palette. The thick black curves mark the photospheric PIL.}
\label{NLFFF}
\end{figure}

The NLFFF extrapolation results are shown in Figures~\ref{NLFFF}(a) and (b), where a set of selected magnetic field lines (violet) is plotted. These field lines are started from the points in the PIL region, where the strong PVEC density (shown by the blue-red base maps) are concentrated. Figures~\ref{NLFFF}(a) and (b) correspond to the very begin and the major peak (i.e. the GOES X-ray flux maximum) of the entire flare, respectively. At the beginning of the flare, the highly sheared intersecting magnetic field lines were involved into the initial energy release process above the PIL. This magnetic configuration is favorable for the TCMR, and the compact twisted magnetic field lines observed at the flare maximum along the PIL in the form of a magnetic flux rope is a result of the magnetic restructuring due to the TCMR process (see also Paper~I). It can be seen that, in general, magnetic field lines around the PIL became more pressed to the solar surface during the flare. (More detailed dynamics of the magnetic field lines in the PIL region is shown in the movie available in the supplementary materials to the paper on the journal website.)

Figures~\ref{NLFFF}(c) and (d) present the 3D structure of coronal electric current density in the PIL region for the two time instants (the same as in Figures~\ref{NLFFF}(a) and (b), respectively). It is shown by the white semitransparent surface corresponding to the coronal electric current density of constant level 55~mA/m$^{2}$ (this value is arbitrary selected and is about 5$\sigma_{j_z}$), which is calculated using the Ampere's law applied to the NLFFF extrapolation results. One can see that the flare energy release lead to expansion of the current-carrying region elongated along the PIL. The pre-flare state was characterized by the thinner channel of strong electric current density. Then this channel became thicker across the PIL. Here we need to note that the NLFFF approximation gives only electric currents flowing along magnetic field lines. (Temporal evolution of the current surface is also presented in the movie which can be found in the supplementary materials to the paper on the journal website.)

The NLFFF modelling reveals the flare-related magnetic field restructuring around the PIL. Interaction of the crossed sheared magnetic loops at the PIL lead to the formation of region at the PIL with the enhanced horizontal magnetic field component when vertical component was quasi stable. It is resulted from formation of the flux-rope-like magnetic structure along the PIL and the sheared magnetic arcade and expansion of the elongated electric current channel in the corona. It also corresponds to formation of the region where we observed the enhancement of the total PVEC (described in the previous section).

%%%%%%%%%%%%%%%%%%%%%%%%%%%%%%%%%%%%%%%%%%%%%%%%%%%%%%%%%%%%%%%%%%%%%%%%%%%%%%%%%%%%%%%%%%%%%%%%%%%%%%%%%%%%%%%%%%%%%%%%%%%%%%%%%%%%%%%%%%%%%%%%%%%%%%%%%%%%%%%%%%%%%%
\section{CHANGES OF MAGNETIC FIELD AND PVEC IN THE UV FLARE RIBBONS}
\label{ribbs}

Flare ribbons are associated with footpoints of magnetic field lines directly connected with the magnetic reconnection site and their observation is important for diagnostic of the flare energy release process. Here we concentrate on a detailed analysis of variations of the magnetic field and PVEC density inside the fare ribbons in the PIL region using 135-second magnetograms.
%Previous researches used mostly LOS magnetograms and vector magnetograms with low temporal resolution (like 720-second HMI vector magnetograms) to study non-stationary magnetic fields in flare regions. But it is not very suitable for highly dynamic solar flares revealing three-dimensional magnetic structure. New HMI 135-second vector magnetograms allow to trace better all magnetic field components and PVEC density distribution in the moving flare ribbons.

To study the dynamics of the flare ribbons we used AIA UV 1700~\AA{} images. This UV emission is generated in the chromosphere. Temporal resolution of this data product is 24 seconds and the pixel size is $\approx 0.6^{\prime\prime}$. We decided to use these images instead of SOT Ca~II images with better spatial resolution mainly due to the limited field-of-view (FOV) of the last one. SOT did not observe some distant flare emission sources, while AIA observes the whole solar disk. Moreover, SOT has only 1-minute temporal resolution when flare observational regime stops. In our case, such resolution was before the flare onset and after the second sub-flare.

\begin{figure}[t]
\centering
\includegraphics[width=1.0\linewidth]{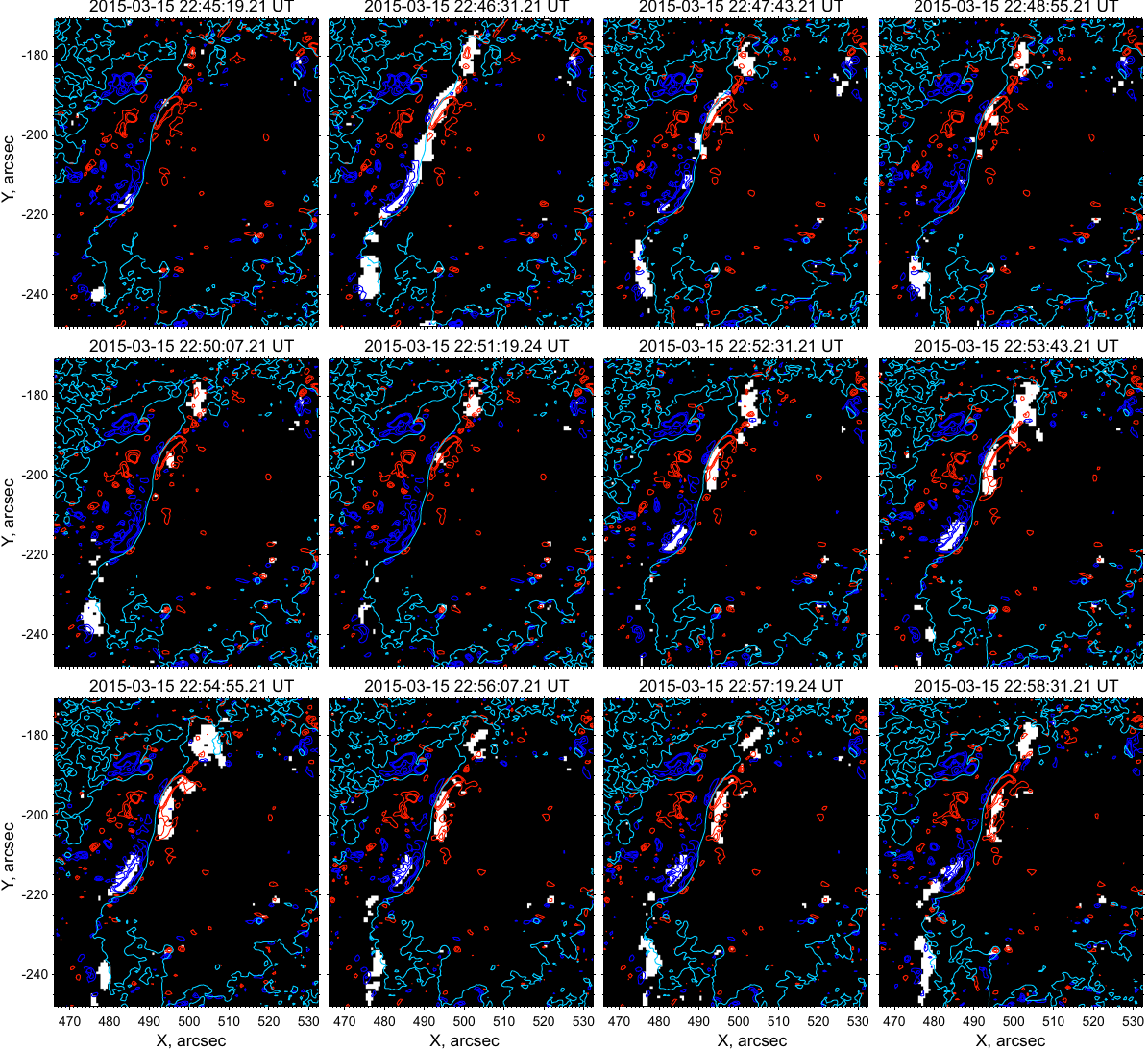}
\caption{Temporal sequence of binary maps (black-white background images) showing the flare ribbon positions deduced from the AIA/SDO UV 1700 \AA{} images. Corresponding times are shown above the panels. Pixels in these maps have only two values: 0 or 1. Value 1 means that this pixel belongs to the flare ribbons. The ribbons were extracted by thresholding images with the threshold value of 3800~DNs. The red and blue contours (23, 39 and 54 mA/m$^2$) correspond to the negative and positive PVEC, respectively. The photospheric PIL is marked by the cyan curves.}
\label{AIArib}
\end{figure}

In each image, the area of the flare ribbons is calculated as a sum of pixels with intensity values higher than the arbitrary selected threshold of 3800~DNs. Figure\,\ref{AIArib} presents temporal sequence of binary maps (the black-white background images) showing the flare ribbon positions deduced from AIA UV 1700 \AA{} images. Each pixel in these maps can have value of 0 (black) or 1 (white). Value 1 means that a given pixel is inside the flare ribbons. Positions of the flare ribbons are compared with the regions of strong PVEC density shown by the red and blue contours corresponding to the negative and positive PVEC density, respectively. In general, the flare ribbons appeared around the PIL in the regions of enhanced PVEC density. However, parts of the ribbons did not coincide with the strong PVECs in some time intervals, in particular, the southern part of the ribbons ($x \approx -240^{\prime\prime}, y \approx 476^{\prime\prime}$) at 22:46:31--22:50:07 UT and 22:54:55--22:58:31 UT.

%The brightest flare ribbons are close to the strongest PVEC density intensification qualitatively. However, there is no exact correspondence. The distant southern and northern UV sources do not have corresponding strong PVEC density. The bottom panels reveal small (a few arcseconds) displacement of the flare ribbons from the PIL and appearance of corresponding PVEC density intensifications. Thus, dynamics of the flare ribbons and regions with enhanced PVEC density are in qualitative matching with each other.

\begin{figure}[t]
\centering
\includegraphics[width=1.0\linewidth]{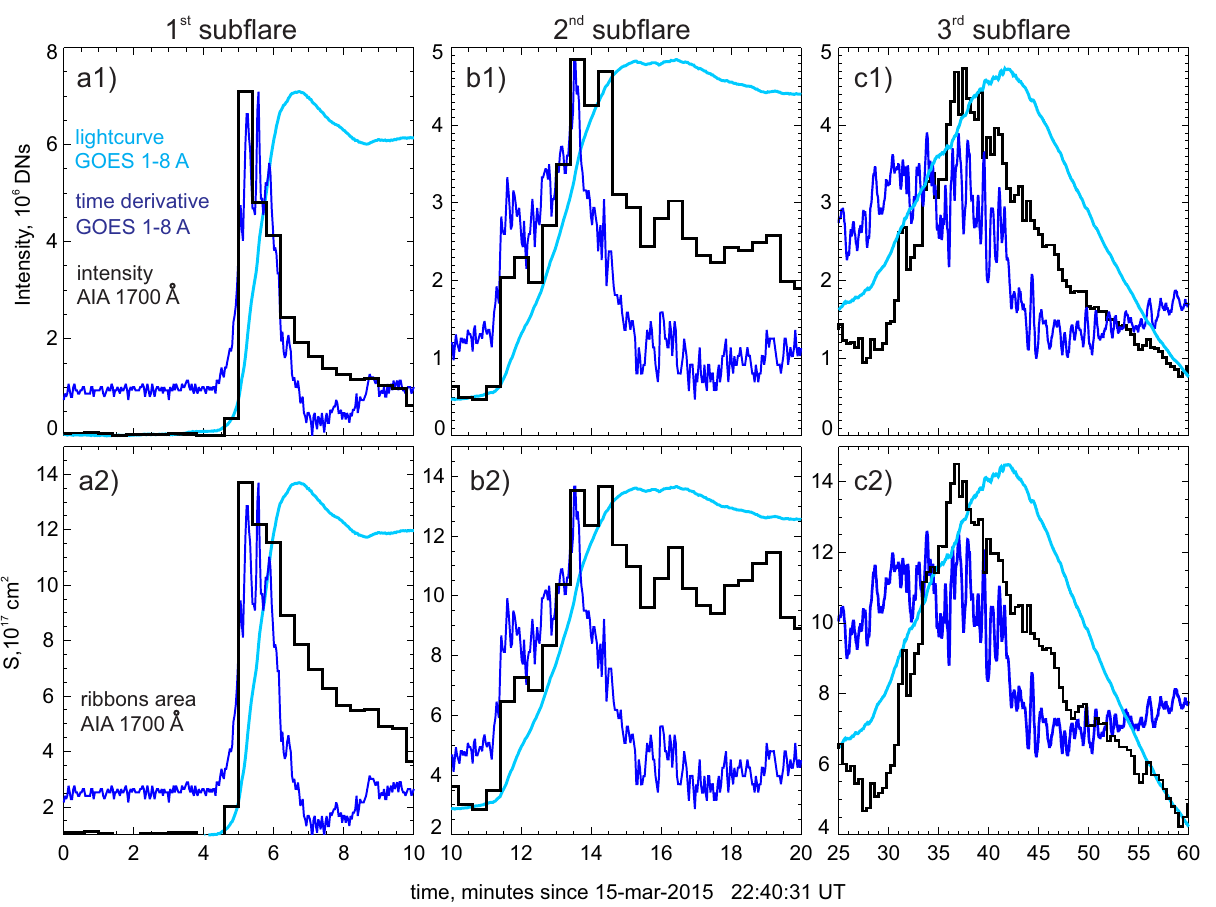}
\caption{Time profiles are plotted for three subflares (columns a-c). The temporal dynamics of the total UV intensity of the flare ribbons is shown in panel (a1-c1) by the black histogram. Temporal dynamics of the flare ribbon area (black histogram) deduced from the AIA UV 1700~\AA{} images (a2-c2). The area is calculated as a sum of pixels with the intensity values higher than the threshold of 3800~DNs (the background images in Fig.\,\ref{AIArib}). The cyan and black lines mark the GOES 1-8~\AA{} lightcurve and its time derivative, respectively.}
\label{AIAribParams}
\end{figure}

In Figure~\ref{AIAribParams} the temporal profiles of the flare ribbon area (a1-c1) and the total UV intensity (a2-c2) are shown by the black histograms and compared with the GOES 1--8~\AA{} lightcurve (cyan) and its time derivative (blue). Three columns of the figure correspond to three subflares. The fastest UV intensity growth and the maximal value was during the first sub-flare with the largest area change of $\approx 1.3 \times 10^{18}$~cm$^2$/s. The maximal total area of the ribbons is about $1.4\times 10^{18}$~cm$^2$.

The sequence of the UV images reveals the sharp enhancement of the ribbon area during the first frames (Figure~\ref{AIAribParams}) of the first sub-flare. Dynamics of the ribbons can be splitted into two phases. The first one (before around 22:49 UT) is not very evident as we do not have sufficient temporal and spatial resolution. However, comparing the first and the second images (two left top images in Figure\,\ref{AIArib}) one can see the fast expansion (or elongation) of the ribbons along the PIL from the initial compact brightenings. The second stage (mainly after around 22:49 UT) is characterized by the gradual separation of the ribbons out from the PIL.

\begin{figure}[t]
\centering
\includegraphics[width=1.0\linewidth]{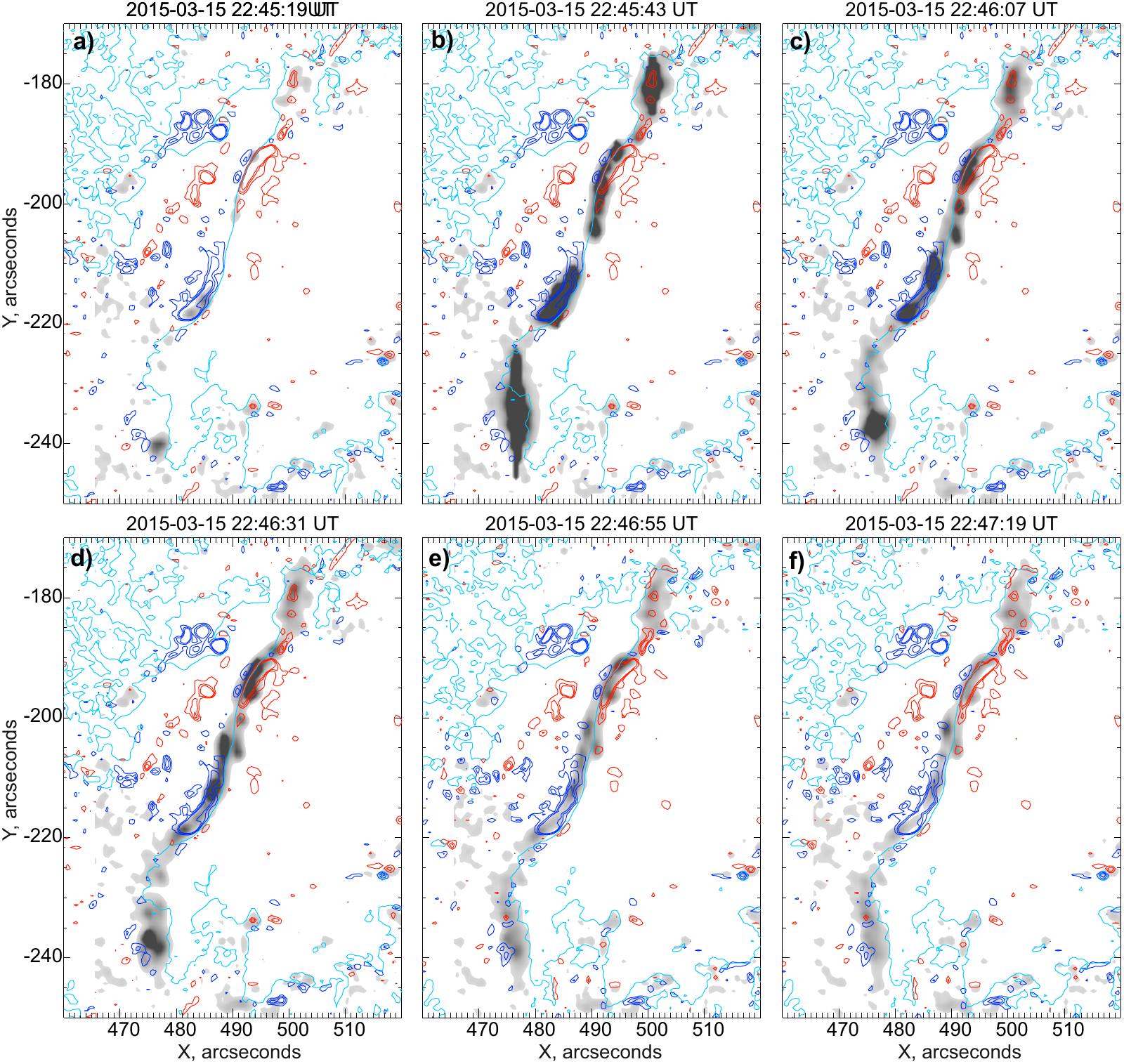}
\caption{Temporal sequence of AIA 1700~\AA{} images (white-black background maps) showing positions of the flare ribbons during the first sub-flare of the 15 March 2015 flare studied. The corresponding time is shown above the panels. The red and blue contours (23, 39 and 54 mA/m$^2$) correspond to the negative and positive PVEC, respectively. The photospheric PIL is marked by the cyan curves.}
\label{AIArib_1st}
\end{figure}

To investigate details of the evolution of the ribbons during the first sub-flare we present Figure~\ref{AIArib_1st} with six (non-binary) UV images compared with PVEC density contour maps. Initially (panel a) we observe the small weak emission sources in the regions of strong PVEC density around the PIL. Then (in panel b), we see the large scale ribbons with non-uniform brightness distribution from $\approx -245^{\prime\prime}$ up to $\approx -175^{\prime\prime}$ along the south-north direction. This expansion occurred during the 24-second time interval of the temporal resolution for AIA UV 1700~\AA{} images. One can also see (panels (c)--(f)) that later the length of the flare ribbons almost did not change, while the ribbon emission intensity varied with time.

The ribbon separation during the first subflare (the distance between the ribbons in a direction perpendicular to the PIL) was about 1~Mm, that means that the emission sources were very close to the PIL and comparable with the ribbons width. Considering the whole event (all three sub-flares), the averaged ribbon velocity ($V_{rib}$) in the perpendicular direction to the PIL is $\approx 5^{\prime\prime}$ per 10 minutes or $V_{\perp} \approx 6.25$~km/s. That's why, during the first subflare the perpendicular displacement of the ribbons was very small $<1^{\prime\prime}$. To estimate expansion rate $V_{\perp}$ during the very beginning of the impulsive phase (the upper limit) of the first sub-flare let's assume fast expansion of the flare ribbon up to its width from the initial brightenings. In this case, the approximate velocity is $\approx 3^{\prime\prime}$ during 24 seconds (the time cadence of AIA 1700~\AA{} images), or $V_{\perp} \approx 100$~km/s. Actual velocity during 24 seconds can be higher or lower than this value within range of $\approx 40-160$~km/s calculated considering $\sigma_t = 12$~seconds (half duration of AIA time resolution) and $\sigma_pix = 0.3^{\prime\prime}$ (half size of AIA pixel).

Let's estimate the parallel velocity $V_{||}$ of the ribbon elongation during the first sub-flare. It cannot be measured directly, because of its fast elongation time relatively to the AIA 1700~A cadence (24~s). However, we can estimate it by the following way. Considering ribbons width $V_{\perp}t$ much smaller then its length $V_{||}t$, the total area $S(t)$ of both flare ribbons can be approximated as $S(t) = 2V_{||}V_{\perp}t^2$ and the expansion rate is $dS/dt = 4V_{||}V_{\perp}t$. Considering the first sub-flare, one can estimate the elongation velocity as $V_{||} = \Delta S/(4V_{\perp} \Delta t^2) \approx 560$~km/s for the beginning of the impulsive phase taking $\Delta t = 24$~seconds, $\Delta S \approx 1.3 \times 10^{18}$~cm$^2$ and $V_{\perp} \approx 100$~km/s (see estimation above). As we considered the maximal possible value of $V_{\perp}$, the obtained $V_{||}$ is the lower limit. If we will take smaller $\Delta t$ than velocity will be larger up to the Alfven velocity $V_{A} \approx  6900$~km/s (upper limit) considering $B = 1000$~Gauss and ion density of $n_{i}=10^{11}$~cm$^{-3}$ (see Paper~I). For $v_{||}=v_A$ the ribbons elongation developed on the time scale $\Delta t\approx 0.5$~s that is much less than actual temporal resolution of AIA data. To sum up, $V_{||}\gg V_{\perp}$ during the beginning of the impulsive phase. Than the flare ribbon elongation was finished and we observed their slow motion out from the PIL. This picture is quite similar to the two step reconnection process discussed in \citep[e.g.][]{Qiu2010,Priest2017}.

%In the case of $V_{\perp} \approx 6.25$~km/s, $V_{||}\approx 9000$~km/s, which is comparable, in order of magnitude, with the Alfven velocity $V_{A} \approx  6900$~km/s considering $B = 1000$~Gauss and ion density of $n_{i}=10^{11}$~cm$^{-3}$ (see Paper~I).

\begin{figure}[t]
\centering
\includegraphics[width=1.0\linewidth]{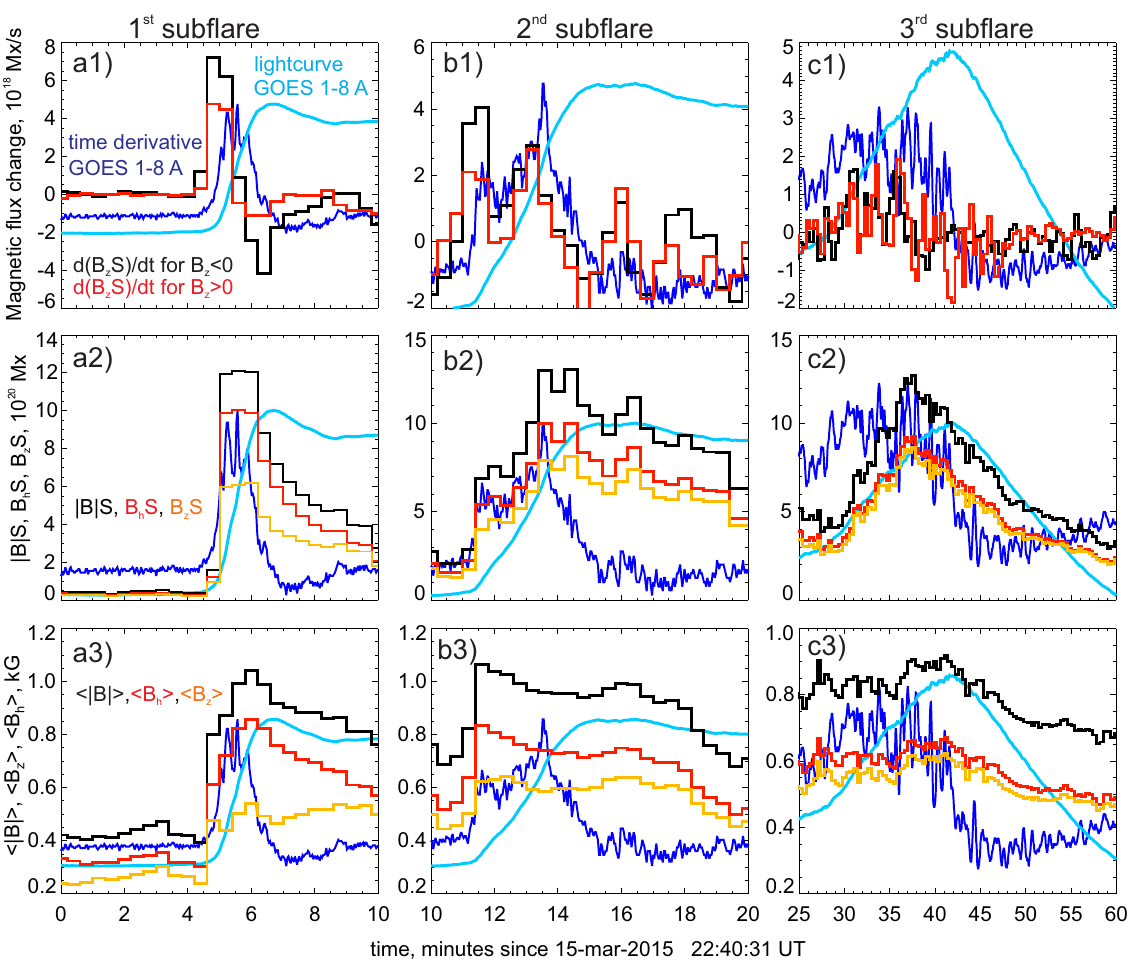}
\caption{Time profiles are plotted for three subflares (columns a-c). (a1-c1) Time derivatives of the (vertical) magnetic flux inside the flare ribbons deduced by the thresholding AIA UV 1700~\AA{} images (Fig.\,\ref{AIArib}). The negative and positive fluxes are marked by the black and red colors, respectively. (a2-c2) Temporal dynamics of the magnetic fluxes, which are calculated for the magnetic field absolute value (black), vertical (orange) and horizonal (red) components. (a3-c3) Time profiles of the average magnetic field values in the flare ribbons. Colors mark different magnetic field components, similar to (a2-c2). The cyan and blue lines mark the GOES 1-8~\AA{} lightcurve and its time derivative, respectively.}
\label{AIAribParams_B}
\end{figure}

Further we will investigate dynamics of the magnetic field and PVEC inside the flare ribbons for all three sub-flares. To calculate the magnetic field parameters inside the flare ribbons at the time of the selected AIA UV 24-second frame we used linear interpolation between corresponding pixels of two neighboring HMI 135-second magnetograms (the UV frame is between these two magnetograms). Time derivatives of the magnetic (vertical component) flux inside the flare ribbons is shown in Figure\ref{AIAribParams_B}(a1-c1) by the black (negative) and red (positive) histograms. Such time derivative is usually referred as the magnetic reconnection rate. The largest reconnection rate of $\approx 7 \times 10^{18}$~Mx/s was achieved in the beginning of the first sub-flare. This reconnection rate is almost two and six times larger than for the second and third sub-flares, respectively. Thus, dynamics of reconnection rate nicely correlates to heating rate of the whole flare. It is also worth noting that two enhancements of the reconnection rate during the second sub-flare are in accordance with two major heating bursts deduced from the time derivative of the GOES 1--8~\AA{} lightcurve.

Figures\,\ref{AIAribParams_B}(a2-c2) and (a3-c3) present variations of the magnetic field components inside the flare ribbons. Panels (a2-c2) demonstrates the magnetic flux (orange) around the PIL, which is calculated using the vertical magnetic field component. This panel also demonstrates the nominal magnetic fluxes deduced for the absolute value of the magnetic field (black) and for the horizontal magnetic field component (red). These nominal magnetic fluxes are introduced to make comparison between the magnetic field components over the total area of the flare ribbons. The time profiles of the magnetic field components averaged over the flare ribbon area are plotted in panels (a3-c3) with the same meaning of colors as in panels (a2-c2). Each sub-flare was characterized by approximately the same enhancement of the nominal total magnetic flux with the magnitude of $\approx 6.5 \times 10^{19}$~Mx. The first and second sub-flares were characterized by the averaged absolute magnetic field value of $\approx 1100$~G, when the third one had slightly less value of $\approx 800$~G. However, the main result is that the magnetic field inside the flare ribbons associated with the first sub-flare were more horizontal than in the case of the subsequent sub-flares. At the time of the maximal $|\vec{B}|$ during the first sub-flare the ratio $<B_{h}>/<B_{z}> \approx 1.62$ and the flux ratio was about 1.66. For the second and third sub-flares these ratios had the following values: 1.3 and 1.2, and 1.2 and 1.1, respectively.

\begin{figure}[t]
\centering
\includegraphics[width=1.0\linewidth]{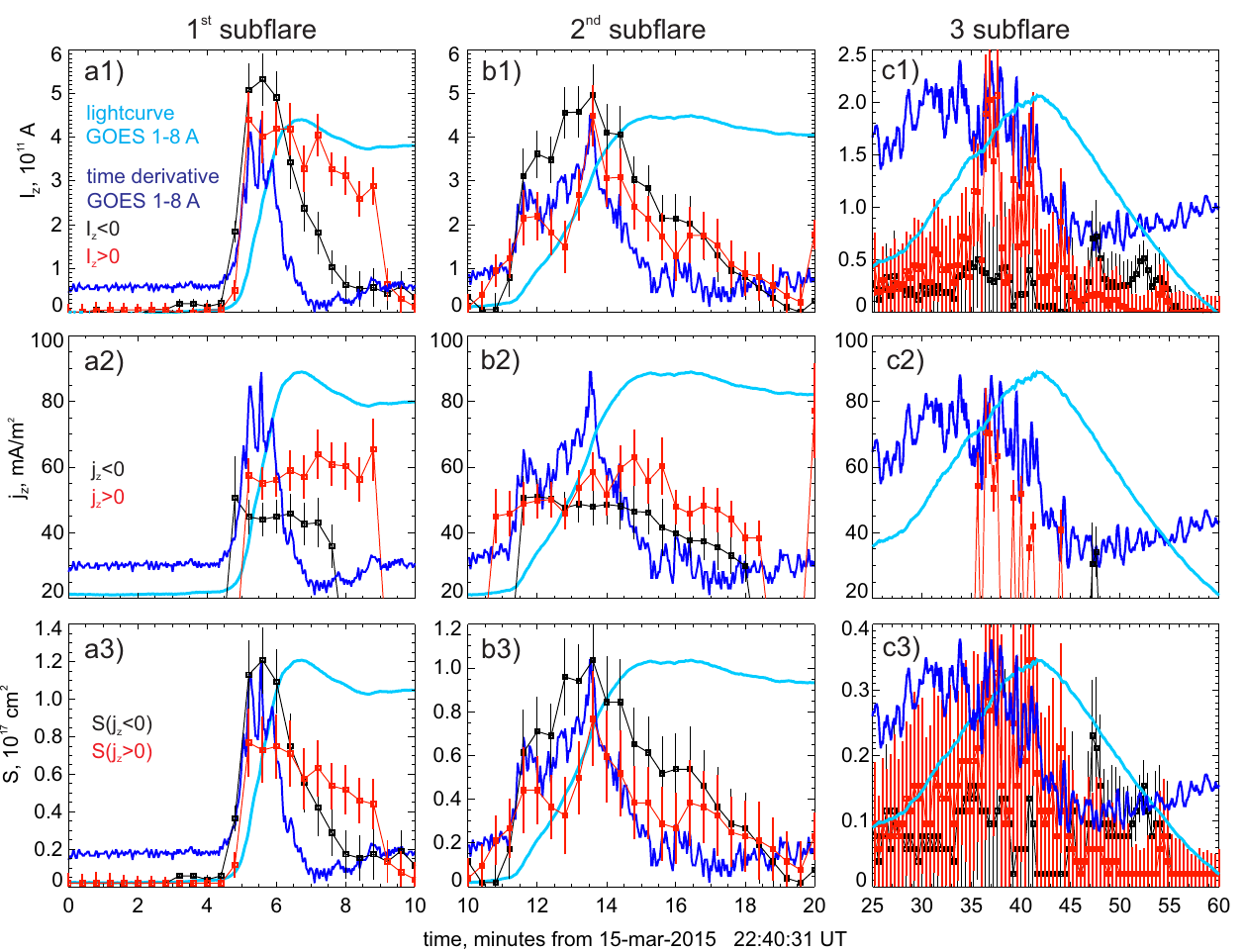}
\caption{Time profiles are plotted for three subflares (columns a-c). (a1-c1) Time profiles of the total PVEC inside the flare ribbons. (a2-c2) Time profiles of the average PVEC density inside the flare ribbons. (a3-c3) Time profiles of the total area characterizing the considered regions with strong PVEC within the flare ribbons. Estimates of the measurement errors are shown by the vertical bars. The red and black colors correspond to positive and negative electric currents, respectively. The cyan and blue lines mark the GOES 1-8~\AA{} lightcurve and its time derivative, respectively.}
\label{AIAribParams_j}
\end{figure}

Figure~\ref{AIAribParams_j} shows the following time profiles: (a1-c1) the total PVEC $I_{z}\left(t\right)$ inside the flare ribbons, calculated as a sum of pixel values above 35~mA/m$^{2} \approx 3\sigma(j_{z})$ and multiplied by the pixel area; (a2-c2) the effective (or averaged) PVEC density $\left\langle j_{z}\left(t\right) \right\rangle$ inside the flare ribbons estimated as the ratio of the total PVEC $I_{z}\left(t\right)$ (panel~(a)) to its area $S\left(t\right)$, whose time profile is shown in panel (a3-c3). The time profiles are shown for the positive (red) and negative (black) PVEC density signs. The total PVEC in the ribbons during the third sub-flare was small comparing with the first and second sub-flares having $I_{z} \approx 4-5 \times 10^{11}$~A. The maximal effective PVEC density was about 60~mA/m$^{2}$ for the first and second bursts. The flare ribbons were partially covered by the regions with strong PVEC density. From panels (a3-c3), the maximal area of the regions with strong ($>35$~mA/m$^{2}$) PVEC density was $\approx 10^{17}$~cm$^2$, which is about 10\,\% from the total area of the flare ribbons. From Figure~\ref{AIAribParams_j} we can make an important conclusion that the temporal variations of plasma heating and electron acceleration efficiency (inferred from the time derivative of the GOES 1--8~\AA{} lightcurve and the Neupert effect) is consistent, in general, with the variations of the effective PVEC density inside the flare ribbons. Although there are some discrepancies, in particular, the peaks of the positive $\left\langle j_{z}\left(t\right) \right\rangle$ of the first and second sub-flares are a bit (2--3 min) longer than the corresponding peaks of the time derivative of the GOES 1--8~\AA{} lightcurve.

The analysis done in this section allowed us to make detailed comparison of the flare energy release with the changes of the magnetic field and PVEC in the flare ribbons, connected with the energy release site. We found good time consistency of the flare energy release efficiency (approximated by the time derivative of the GOES 1--8~\AA{} lightcurve) with the magnetic reconnection rate and the total PVEC inside the ribbons.

%%%%%%%%%%%%%%%%%%%%%%%%%%%%%%%%%%%%%%%%%%%%%%%%%%%%%%%%%%%%%%%%%%%%%%%%%%%%%%%%%%%%%%%%%%%%%%%%%%%%%%%%%%%%%%%%%%%%%%%%%%%%%%%%%%%%%%%%%%%%%%%%%%%%%%%%%%%%%%%%%%%%%%
\section{DISCUSSION}

\subsection{MAIN RESULTS}
\label{MainRes}

Using the new HMI vector magnetograms with the high time cadence of 135~s, we investigated dynamics of the photospheric magnetic field and PVEC in the PIL region during the three successive sub-flares of the confined M1.2 class solar flare on 2015 March 15 (SOL2015-03-15T22:43), previously studied in Paper~I \citep{Sharykin2018}. The following main data analysis results were obtained:

\begin{itemize}

\item {The flare X-ray sources, as well as the optical and UV flare ribbons, were located, in general, in the regions of the strong horizontal photospheric magnetic field, PVEC density and the horizontal gradient of the vertical magnetic field component on the photosphere around the PIL.}

\item {The photospheric magnetic field was mostly horizontal near the PIL region. The magnetic field in the flare region} became more horizontal during the flare. The large fraction (more than half) of the magnetic flux was concentrated in the strong ($B>1000$~G) magnetic field. There were no fast flare-associated changes of the vertical magnetic field component.

\item {The total PVEC near the PIL region increased simultaneously with the increasing photospheric cross-sectional area of the current-carrying magnetic structure. The effective PVEC density first increased impulsively with the flare onset and then decreased gradually with the flare development. Expansion of the regions with the strong PVEC is in qualitative accordance with the motion of the flare ribbons from the PIL.}

\item {The flare ribbons penetrated into the regions with more vertical magnetic field in course of the flare, while they were initially characterized by the dominant horizontal magnetic component. The expansion of the flare ribbons was characterized by enhancement of the total PVEC and effective PVEC density inside them, which had tendency to coincide temporally with the lightcurves of microwave emission and the time derivative of the SXR lightcurve.}

\item {The NLFFF extrapolation based on the HMI 135-sec magnetograms showed that (a) the pre-flare state was characterized by the highly sheared magnetic structure intersecting above the PIL favorable for the three-dimensional (3D) TCMR; (b) the magnetic field lines started from the regions of strong PVEC became shorter and formed the low-lying magnetic flux-rope-like structure embedded in the sheared magnetic arcade along the PIL at the end of the flare. }

\end{itemize}

In the following subsections we will discuss these observational phenomena jointly with the results from Paper~I in the context of the TCMR above the photospheric PIL. We will try to give qualitative interpretation of the phenomena observed during the flare and make the order of magnitude estimations of some important physical parameters related to the energy release processes and electron acceleration.

%The energetics and efficiency of the magnetic reconnection, filamentation of energy release site, dynamics of the magnetic fields and electric currents, acceleration of electrons will be discussed in the separate subsections.

%---------------------------------------------------------------------------------------------------------------------------------------------------------------------

\subsection{DYNAMICS OF MAGNETIC FIELD AND ELECTRIC CURRENTS IN THE FLARE REGION}

%One of the main tasks of this work is to investigate dynamics of the photospheric magnetic field components and PVEC in the vicinity of the PIL where the flare energy release developed in the conditions of three-dimensional restructuring of the magnetic field lines within the found sheared magnetic structure. Using the HMI 135-second vector magnetograms we found the increase of the horizontal magnetic component near the PIL during the flare. Similar behavior of the horizontal magnetic component during flares was shown in other works \citep[e.g.][]{Wang1994,Sun2012,Wang2012,Petrie2013,Sun2017,Lu2018}. From the ``standard'' eruptive flare model it follows that magnetic field can become more vertical around the PIL due to removal of a portion of horizontal magnetic flux by an erupting flux rope. However, due to contraction (collapse) of magnetic loops reconnected underneath an erupting flux rope \citep{Somov1997,Hudson2000}, magnetic field can become more horizontal near the PIL. In the case of the TCMR process \citep{Moore1992,Moore2001} during a confined flare, like in the event studied here, the sheared magnetic loops interact and reconnect above the PIL, which results in formation of a small magnetic arcade moving/collapsing towards the photosphere. Thus, we observe the growth of the horizontal magnetic component near the PIL during the flare.

Here we will discuss dynamics of the photospheric magnetic field components and PVEC (see the second and third items in the list of the main results in Section\,\ref{MainRes}) in the vicinity of the PIL where the flare energy release developed in the conditions of 3D restructuring of the magnetic field lines within the found sheared magnetic structure.

The found increase of the horizontal magnetic component near the PIL during the flare can be explained in the frame of the TCMR scenario. The sheared magnetic loops interact and reconnect above the PIL, which results in formation of a small magnetic arcade moving/collapsing towards the photosphere. That is why the growth of the horizontal magnetic component near the PIL is observed during the flare.

The larger and higher magnetic field lines are also formed due to this reconnection process. These field lines move upwards and interact with the overlying magnetic field above the primary energy release site. Subsequent episodes of magnetic reconnection are triggered in this case. We observe it, firstly, as the initial fast elongation of the flare ribbons along the PIL, and after that as the small displacement of the flare ribbons out from the PIL and appearance of the expanding hot sheared magnetic arcade (Figure\,\ref{preflare}(b)). The velocity of this expansion is rather small that is in agreement with the ``slipping'' reconnection in the confined flares \citep[][]{Hinterreiter2018}. In other words the expansion is a result of involvement of new magnetic loops lying above the already reconnected ones. The reconnection happens between higher and higher, more distant magnetic loops from the PIL, which are less sheared and in a more potential state \citep{Priest2017,Qiu2017}. If a full eruption does not occur, then this process stops at some point (at some height). Determining the reason for stopping this process and a lack of developed eruption in the flare studied is beyond the scope of this work \citep[see, e.g.,][]{Amari2018}.

Let's now discuss the observed dynamics of the PVEC in the flare studied (see the third and forth items in the list of the main results in Section\,\ref{MainRes}). First of all, we need to mention that the observed dynamics is not related to the projection effect. The thing is that the observed enhancement of the total PVEC in the flare region may be connected with the magnetic field verticalization near the PIL. However, the magnetic field near the PIL become more horizontal during the flare and, thus, the increase of PVEC cannot be the result of magnetic field inclination change at the foot of the flare loops. Our working hypothesis is that the detected changes of the PVEC density could be connected with the electric current generation/amplification in the current sheet(s) above the PIL in course of the 3D TCMR. Indeed, in contrast to the 2D ``standard'' model, in the 3D magnetic configuration, the current flowing in the coronal current sheet(s) must be closed somewhere. We assume that this may take place, at least partially, under the photosphere. Thus, the appearance/amplification of the current in the corona can be accompanied, possibly with some delays, by the appearance/amplification of vertical currents on the photosphere. This is ideologically similar to the results obtained by \citet{Janvier2014,Janvier2016} for two powerful eruptive X-class flares on 2011 February 15 and 2011 September 6 \citep[see also][ for discussions]{Schmieder2018}.

As we discussed above, the magnetic reconnection in the flare studied is a non-stationary process involving different magnetic loops of various spatial scales. To analyze dynamics of electric current in the current sheet qualitatively, firstly, let's consider a single reconnection episode and try to understand the situation when we have a pulse of electric current density and gradual increase of the total electric current. For simplicity, we will consider a standard rectangular diffusion region with height $l$ and width $\delta$ (Fig.\,\ref{scheme1}b). Magnetic field at the boundaries of this region has three components: the guide ($B_{||}$; along the PIL),  perpendicular ($B_{\perp}=gB_{in}$; i.e. vertical), and transverse $B_{x}$ (i.e. across the sheet) components. Here $g$ is a geometric factor connected with the magnetic shear across the current sheet. It is difficult to estimate $g$ from observations, and we just assume that it is less than unity. To estimate the total electric current through the current sheet we use the Ampere's law in the integral form:
\begin{equation}
J_{||} = \int_{S_C} j_{||}dS = \frac{c}{4\pi}\oint_C\vec{B}\cdot\vec{ds} = \frac{c}{2\pi}B_{\perp} l \left[1+\frac{B_{x}}{B_{\perp}}\frac{\delta}{l}\right] \approx \frac{c}{2\pi}B_{\perp}l
\label{eq:JII}
\end{equation}
Here $S_{C}=\delta l$ is a cross-sectional area of the current sheet and $C$ means the contour marking the boundary of this area. This approximate formula for $J_{||}$ is derived assuming $B_{x}\delta/B_{\perp}l \ll 1$ that is quite natural considering magnetic flux conservation $v_{in}B_{\perp}=v_{A}B_{x}$. Thus, $j_{||} \propto B_{\perp}/\delta$ and $J_{||} \propto B_{\perp}l$. The time derivatives of the electric current density and the total electric current are $j_{||t} \propto B_{\perp t}/\delta - B_{\perp}\delta_t/\delta^2$ and $J_{||t} \propto lB_{\perp t} + B_{\perp}l_t$, respectively, where subscript $t$ means time derivative. Thus, to achieve $j_{||t} >0$ and $J_{||t} >0$ one has to guarantee simultaneous fulfillment of the following conditions: $B_{\perp t}/B_{\perp}>\delta_{t}/\delta$ and $B_{\perp t}/B_{\perp}>-l_{t}/l$. In other words, relative growth rate of the magnetic field near the diffusion region boundary should be larger than change of the current sheet width and the change of the height with opposite sign. The obvious way to fulfill these inequalities is to assume the current sheet thinning, elongation and enhancement of $B_{\perp}$ value at the reconnection region boundary. Such dynamics of a current sheet was simulated for eruptive solar flares and discussed by \cite{Janvier2016}. We suggest that the similar behavior for the 3D magnetic reconnection with a guide field in the closed magnetic configuration like in the frame of the 3D TCMR scenario considered here is also possible. To achieve $B_{\perp t}>0$ we can suggest the induction equation $B_{\perp t} = [v_{in}B_{||}]_x>0$. It means that the inflow across the current sheet leads to enhancement of the perpendicular magnetic field.

After the pulse of $<j_{z}>$, we observed the situation when the total electric current and current density show different trends: $j_{||t} <0$ and $J_{||t} >0$ (see Figure\,\ref{Idyn}(e1--e3)). It can described by the following inequalities: $B_{\perp t}/B_{\perp}<\delta_t/\delta$ and $B_{\perp t}/B_{\perp}>-l_t/l$. We suggest that the most reliable and simple scenario explaining these inequalities is the magnetic annihilation process without magnetic advection described by the simple equation for the diffusion region: $B_{\perp t} = \mu B_{\perp xx}$, where $\mu$ is the magnetic diffusivity and $xx$ means double $x$ derivative. Considering $B_{\perp t} \approx 0$ at the boundary as there is no advection, we will obtain $\delta_t \propto \sqrt{\mu/t}>0$ and $j_{||t}<0$.

Above we discussed the observed dynamics of $j_{||}$ and $J_{||}$ in the frame of a single reconnection episode. But enhancement of $I_{z}$ on the photosphere and the area with strong PVEC density could be a result of successive involvement of new magnetic field lines, located higher than previously reconnected ones, into the reconnection process. In other words, the flare process may consist of a large set of reconnection episodes, probably in different locations in the corona above the PIL \citep[e.g. ][]{Zimovets2018}. The moving flare ribbons trace magnetic flux associated with magnetic field lines passing through the reconnection region(s). Analysis of high cadence HMI vector magnetograms revealed expansion of the regions with strong PVEC density associated with the moving flare ribbons seen in the UV images. It confirms the idea that the total PVEC in the flare region has tendency to grow due to successive involvement of new magnetic field lines in the reconnection process in the corona. Good time matching between the flare energy release efficiency (observed as the time derivative of the SXR emission lightcurve) and electron acceleration (observed as the HXR and microwave light curves) from one side, and the flare ribbon area, total PVEC and average PVEC density through the ribbons during three subsequent sub-flares (see Figure\,\ref{AIAribParams_j}) from another side, confirms additionally that the flare energy release is associated with the involvement of new current-carrying magnetic elements and local amplification of PVEC near the PIL. 

%This should be related to the spatio-temporal dynamics of the current sheet(s) above the PIL.

Here it is appropriate to note a certain analogy of the observed phenomena with the processes in the Earth's magnetosphere during substorms. It is well known that substorms are the result of magnetic reconnection in the near-Earth magnetotail current sheet \citep[see for review, e.g.,][]{Baker1996,Angelopoulos2008}. This phenomenon is associated with plasma heating and acceleration, generation of earthward and tailward plasma flows accompanied by the magnetic field enhancement, so called dipolarization. These plasma flows distort magnetic field lines that leads to generation of magnetic shear and the field-aligned currents as a consequence \citep[see e.g.][and references therein]{Kepko2015}. Recent observations in the magnetosphere and numerical simulations confirm the close link between plasma outflows from the near-Earth magnetotail reconnection region and generation of the field-aligned currents \citep[e.g.][]{Artemyev2018}. These results indicate ``that the dominant role of the near-Earth magnetotail reconnection in the field-aligned current generation is likely responsible for their transient nature''. The situation could be ideologically similar in the flare investigated in the present work. The active region around the PIL already contained significant quasi-steady electric currents before the flare (and after it), probably generated as a result of long-lasting horizontal movements of the foot of the magnetic loops. The transient amplifications of the vertical currents on the photosphere during the energy release in the sub-flares could be associated with plasma outflows and magnetic field enhancement as the result of magnetic reconnection episodes in the current sheet(s) above the PIL. Further research should show how justified such an analogy is.

\subsection{MAGNETIC RECONNECTION IN THE PIL REGION}

%We consider the confined flare without a fully-developed eruption and CME. Analysis of the magnetic field extrapolation and morphology of the emission sources in the various wavelength ranges revealed that the flare energy release was developed in the low lying magnetic field structure elongated along the PIL. The presence of the sheared core magnetic structure with crossed magnetic field lines above the PIL is in favor of interaction of the magnetic flux tubes (loops) in the frame of the TCMR scenario \citep{Moore1992,Moore2001}. This process is characterized by strong longtitudal (guide) component of the magnetic field in a reconnecting current sheet. Photospheric vector magnetograms revealed that the flare ribbons were localized in the strong magnetic field with the dominating horizontal component. Possibly, the flare energy release was triggered in the localized magnetic shear layer above the PIL. In this section below we will discuss this 3D magnetic reconnection as the source of energy release of the flare studied and will perform some basic estimations.

\begin{figure}[t]
\centering
\includegraphics[width=1.0\linewidth]{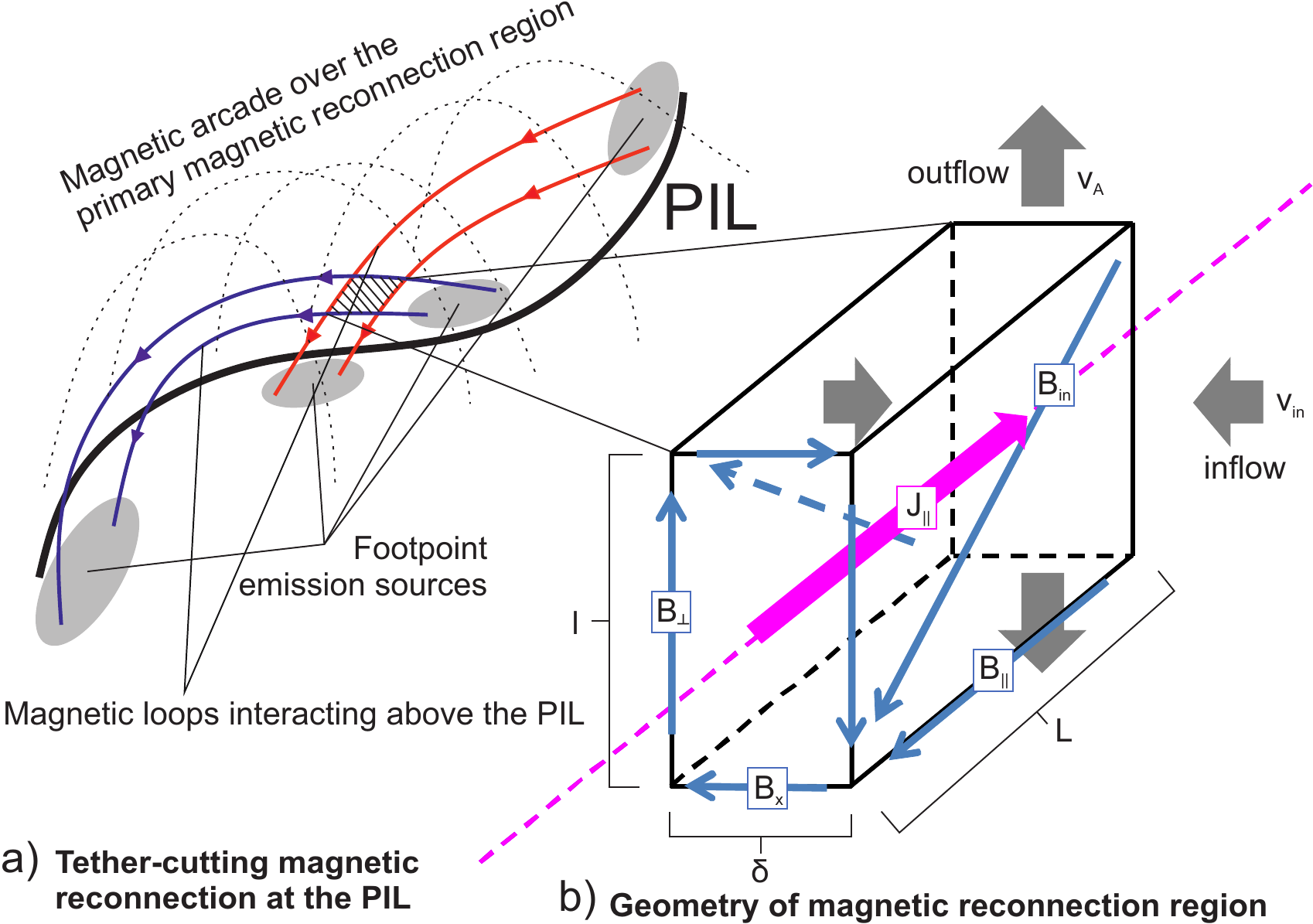}
\caption{This scheme presents geometry of the magnetic reconnection region discussed. Panel~(a) describes the general magnetic field topology in the frame of the TCMR scenario. Panel~(b) shows the magnetic field structure and geometry of the magnetic reconnection region.}
\label{scheme1}
\end{figure}

In  Figure~\ref{scheme1} we present the proposed geometry of the magnetic reconnection region in the frame of the 3D TCMR scenario (panel~(a)). The geometric sizes and magnetic structure of the reconnection site are shown in panel~(a1-c1). The basic parameter of the magnetic reconnection is the reconnection rate $d\phi/dt$, defined as a time derivative of magnetic flux through the flare ribbons \citep[e.g.][]{Forbes2000}. This flux determines magnetic inflow into a current sheet where magnetic reconnection develops. The reconnection rate is proportional to the electric field $E$ in the reconnecting current sheet: $d\phi/dt \sim cEL$, where $c$ is the speed of light, $L$ is the length scale of the current sheet \citep[e.g.][]{Forbes2000}. The maximal value $d\phi/dt \approx 7 \times 10^{18}$~Mx/s and, thus, $E \sim 2/L$~statvolt/cm inside the current sheet with $L$ in Mm. Using the results of Paper~I we take $L \sim 1$~Mm, which is the width of the hot channel at the PIL seen in the ``hot'' EUV bands (AIA/SDO 94 and 131~\AA{}; see Figure~\ref{preflare}). The magnetic reconnection rate can be aslo calculated as $d\phi/dt \sim v_{in} g B_{in} L$, where $v_{in}$ is a velocity of plasma flowing into the current sheet with the length scale $L$ (Figure\,\ref{scheme1}b). It means that the parallel electric field along the sheet is connected with $gB_{in}$ at the current sheet boundaries. It is difficult to estimate the current sheet length $L$. The possible assumption is $L \sim l$, where $l$ is the current sheet height in the vertical direction out from the photosphere and it is about $1$~Mm (i.e. a cross section size of the flare loops; see Paper~I). However, it seems that due to the high shear, $L$ should be a few times larger than $l$. For further estimations let's take in mind that $gL \sim 1$, for $L$ described in Mm. But, not to loss generality, this factor will be written in the following expressions.

We found that $v_{in} \approx 700/(gL_{[Mm]})$~km/s, and the dimensionless reconnection rate (i.e. the Alfven Mach number in the inflow region) $M_{A} = v_{in}/v_{A} \approx 0.1/(gL_{[Mm]})$ for the Alfven velocity $v_{A} = 6900$~km/s considering $B = 1000$~G and ion density of $10^{11}$~cm$^{-3}$. The estimated value of $M_{A}$ is close to the upper limit of the magnetic reconnection rate found in other works \citep[e.g.][]{Yokoyama2001,Isobe2005,Lin2005,Narukage2006,Takasao2012,Su2013,Nishizuka2015,Cheng2018}. The characteristic inflow velocities in that works were in the range of $10-100$~km/s that is significantly smaller relative to our estimation of $v_{in} \approx 700/(gL_{[Mm]})$~km/s. From our point of view, it is unlikely that the driver of the flare energy release in this non-eruptive flare was the large scale displacement of magnetic loops with this velocity. Possibly, such velocity could arise locally within the current sheet due to motion of the separate magnetic elements across the current sheet. For example, formation of magnetic islands (their three-dimensional analog), their relative motion and acceleration up to the Alfven velocity within the current layer will lead to subsequent acceleration of magnetized plasma incoming to the space between magnetic islands \citep[e.g. plasmoid induced magnetic reconnection, ][]{Shibata2001}.

Let's estimate the current sheet width $\delta$. It is reasonable to use the continuity equation for the steady state (reconnection rate maxima when $d^2\phi/dt^2\sim 0$):

\begin{equation}
v_{in} n_{in} l = v_{A} n_{0} \delta
\label{eqcont}
\end{equation}

Thus, $\delta = l (n_{in}/n_{0})M_{A}$ is determined by the plasma compression ratio $n_{in}/n_{0}$ and dimensionless reconnection rate $M_{A}$. From our estimations $M_{A} \approx 0.1/(gL)$. To estimate the compression ratio we assume an incompressible limit as the simplest case. Another option is to assume strong compression from the initial (background) coronal density $n_{in} \sim 10^{9}$~cm$^{-3}$ to the flare super-hot plasma density $n_{0} \approx 10^{11}$~cm$^{-3}$ (see Paper~I). As a result, we obtain $\delta \sim (10^{-3} - 10^{-1})/(gL)$~Mm.

Another important characteristics of the magnetic reconnection process is the ratio $\lambda_{mfp}/L$, where $\lambda_{mfp}$ is the plasma collisional mean free path and $L$ is the characteristic length scale (in our case we assume it to be of the current sheet size). The estimation gives $\lambda_{mfp}\approx 0.5$~Mm for the super-hot plasma temperature $T=40$~MK and density $n\approx 10^{11}$~cm$^{-3}$. Thus, we have collisionless conditions taking the current sheet width $\delta \sim (10^{-3} - 10^{-1})/(gL)$~Mm and height $l\sim 0.5$~Mm. However, for electrons propagating along the current we consider a weakly collisional regime ($\lambda_{mfp}/L=\lambda_{mfp}g/l\gtrsim 1$). It is likely that the magnetic reconnection process was collisionless due to the high plasma temperature.

Let's also estimate the energy release rate in the current sheet with the height of $l=1$~Mm (the width of the ``hot'' EUV channel, see Paper~I for details) and incoming magnetic field $B_{in}=1000$~G using the formula \citep[e.g.][]{Aschwanden2004}:
\begin{equation}
\frac{dE}{dt} = 2 \frac{gB_{in}^{2}}{4\pi}v_{in}Ll = \frac{gB_{in}l}{2\pi}\cdot \frac{d\phi}{dt} \sim 10^{29}g,
\label{eq:dedt}
\end{equation}
ergs/s.

To estimate the inflow energy we neglected the kinetic and thermal energy as the magnetic energy is dominant. During the impulsive phase of the first sub-flare of duration $\Delta t \approx 100$~s the total magnetic energy release is $(dE/dt)\Delta t \sim 10^{31}gL$~ergs that is one order of magnitude larger than the nonthermal particle energy ()$10^{30}$~ergs) and a few times smaller than the change of the free magnetic energy $\approx 2.9-4.6\times 10^{31}$~ergs bearing in mind that $gL\sim 1$ (see Paper~I). Thus, the estimated magnetic reconnection energy release rate does not contradict to the found flare energetics deduced from different emissions. The difference between total magnetic free energy and dissipated magnetic energy in the current sheet can be connected with those fact that we did not take into account subsequent reconnection episodes (the second and the third subflares).

To sum up, despite the fact that the magnetic reconnection during the studied flare has quite high rate (up to $d\phi/dt \approx 7\times 10^{18}$~Mx/s), its dimensionless value ($M_{A} \approx 0.1/(gL)$) is comparable with the upper limit found in other works, where the current sheets located higher in the corona in the eruptive events considered.

%---------------------------------------------------------------------------------------------------------------------------------------------------------------------
\subsection{CURRENT SHEET STABILITY AND FILAMENTATION}

%\subsection{CURRENT SHEET FILAMENTATION AND CONFINEMENT OF THERMAL AND NONTHERMAL ELECTRONS}

% This thin magnetic loop formed in the region of the crossed magnetic loops intersecting in the PIL. Thus, possible current sheet in this geometry has a width limited by sizes of interacting magnetic structures with strong longtitudal magnetic component. In this case we can state that the width of a current sheet is also approximately of 0.5~Mm and $v_{in}\approx 1400 km/s$ with Alfven Mach number $M_A = 0.64$.

The fast inflow velocity ($v_{in} \approx 700/(gL)$~km/s) leading to the large reconnection rate is probably not resulted from the fast macroscopic motions, because we did not found them in the flare region. It is likely to be connected with some local processes around the current sheet. Thus, to trigger the fast magnetic reconnection one need to assume the current sheet achieving unstable conditions somehow. This type of magnetic reconnection is usually called the spontaneous reconnection \cite[e.g.][]{Priest2000}. In the case of the spontaneous reconnection, it is necessary that the current sheet is formed and accumulated sufficient amount of energy to be released during a flare.

% There are a lot of ways to make current sheet unstable, e.g. small perturbations associated with waves or sudden non-periodic slight motions.

\begin{figure}[t]
\centering
\includegraphics[width=0.8\linewidth]{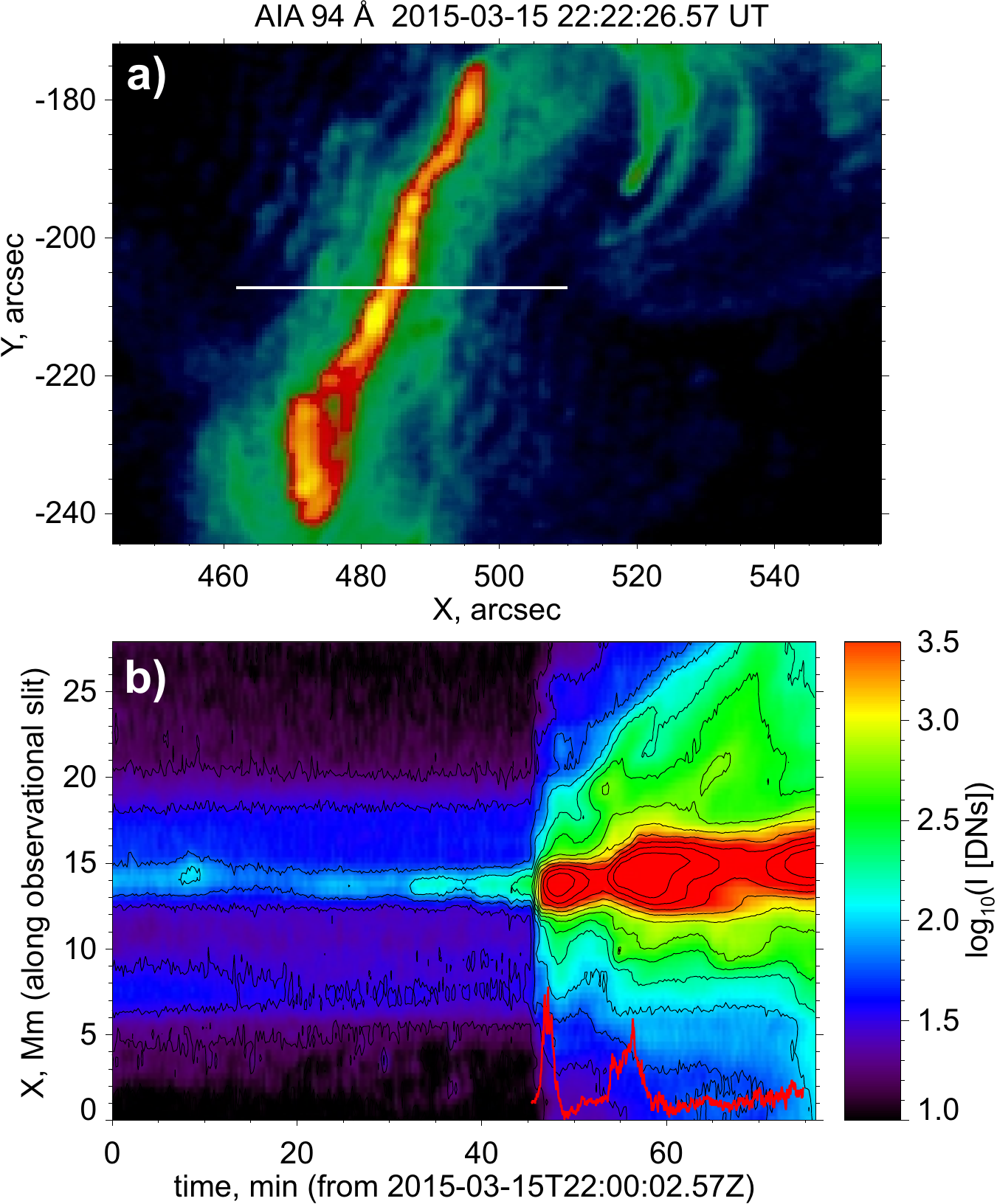}
\caption{(a) The UV AIA 94\,\AA{} image of the active region made around 20 min before the flare impulsive phase onset, showing the presence of the bright elongated source above the PIL. The white horizontal line marks the slit used to make the time-distance plot, shown on (b), to trace the dynamics of the flare energy release around the PIL. To show the time-distance plot with more contrast we overplotted it with the black contours. Both color plots on (a) and (b) are made on a logarithmic scale of intensities. Red curve in the right-bottom part of panel (b) marks the time derivative of the GOES 1-8 \AA{} lightcurve to show the flare energy release rate. Note that time is on a linear scale on the horizontal axis.}
\label{preflare}
\end{figure}

In Figure\,\ref{preflare}(b) we present the time-distance plot for the observational slit (the white horizontal line in panel~(a) with the AIA 94~\AA{} image for one time instant $\approx 20$~min before the flare onset) crossing the bright EUV channel at the PIL. From Figure\,\ref{preflare}(b) we conclude that the energy release (heating) in the bright ``hot'' source elongated along the PIL was observed long before the flare impulsive phase onset (at least before 45~min in Figure\,\ref{preflare}(b)). A similar picture was previously observed before many flares \citep[][and references therein]{Cheng2017}. Sudden enhancements of the channel brightness were detected episodically in the pre-impulsive flare phase. The flare initiation was preceded by the gradual increase of the EUV luminosity for $\approx 10$~min. It seems that the energy release site was already prepared for the flare onset, and once it reached some special conditions for an instability the flare started. The current sheet could exist in the quasi-stationary state and the magnetic reconnection was slow, then the reconnection rate started to grow suddenly, which led to the beginning of the flare and its development. This is consistent with the discussion of the TCMR flare scenario by \citet{Moore1992}.

It is difficult to find a reason for the spontaneous equilibrium loss of the current sheet because of the limited capabilities of the available observational data. However, one of the possible ways to trigger transition from the slow to fast regime of reconnection is to assume the tearing instability leading to formation of current filamentation inside the current sheet. The reason to consider this scenario is the fact that the accelerated electrons and hot plasma were localized in the thin magnetic filament (described in Paper~I). We assume that such filament is a bundle of magnetic flux tubes formed due to the tearing instability inside a non-neutralized current sheet. It was shown \citep[e.g.][]{Kliem1994} that magnetic islands (filaments in 3D) in the current sheet lead to efficient trapping of electrons inside the filaments.

As we know from the theory \citep[e.g. see][]{Priest2014}, the characteristic time $\tau_{tear}$ of the tearing instability is the shortest for large scale perturbations with $k\delta \sim 1$ ($k$ is the wavenumber of perturbation). This time can be estimated as $\sqrt{\tau_{d}\tau_{A}}$ \citep[e.g. subsection 10.2.1 in][]{Aschwanden2004}, where $\tau_A$ is the Alfven time and $\tau_{d} = \delta^{2}/\mu$ is the diffusion time across a current sheet with the width $\delta$ and magnetic diffusivity $\mu=c^{2}/(4\pi\sigma)$, where $\sigma$ is the electrical conductivity. Considering $\delta \sim (10^{-3} - 10^{-1})/(gL)$~Mm (estimated above), one can deduce $\sigma_{eff} = (\tau_{tear}c)^{2} v_{A}/(\delta^{3}4\pi) \sim (2.5\times 10^{11}-2.5\times 10^{17})(gL)^3$, when the classic electrical conductivity is $\sigma_{Sp}\approx 4.8\times 10^{17}$ for $T=40$~MK. Thus, even the classical conductivity could explain formation of magnetic islands in the more narrow current sheet. But in the case of a thicker current sheet one should consider the suppressed (anomalous) electrical conductivity which can be five orders less than the classical one. The electric conductivity reduction can arise due to the presence of turbulence.

%classical Spitzer conductivity gives us unreliable values of $\tau_{tear}$ which should be approximately equal to impulsive phase duration with $\tau\sim$100~seconds. Assuming this time scale one can estimate possible value of

Another reason to suggest the presence of anomalous transport connected with turbulence is the appearance of the super-hot plasma (for details see Paper~I). Confinement of the super-hot plasma on a time scale of the impulsive phase could be due to anomalously slow heat losses from the heated region. Let's estimate characteristic cooling time via heat conduction as $\tau_{cond} \approx 4nk_{B}L^{2}/(kT^{5/2}) \approx 0.13$~s for $L=5$~Mm (half length of 10~Mm magnetic loop) and $T = 40$~MK. Here $k_{B}$ is the Boltzmann constant and $k$ is the thermal Spitzer conductivity coefficient. If we consider saturation limit of heat conduction, when electrons with their thermal velocity transfer energy, we find $\tau_{cond} \sim 1.5V_{Te}L \approx 0.2$~s, where $V_{Te}$ is the velocity of thermal electrons. The values obtained are too small comparing with the duration of the impulsive phase $\sim 100$~s. To resolve this discrepancy, we assume the suppressed heat conduction coefficient by four orders. Some discussion of such possibility, related to the generation of turbulence by beams of accelerated particles, can be found in \cite[e.g.][]{Sharykin2015b}. More detailed discussion of the possible physical reasons for the suppressed heat conduction is out of the scope of the present work.

\subsection{ACCELERATION OF ELECTRONS}

%There are a few main ways to accelerate electrons: large-scale direct electric field; stochastic acceleration by small-scale time-varying electric field; shock acceleration. The last two mechanisms are very difficult for study due to observational limitations. We want to discuss large-scale electric field within a current sheet as a reason for electrons acceleration in the flare studied.

As we know from Paper~I, electrons were accelerated up to the kinetic energy $K_{0} \sim 0.1$~MeV in the first sub-flare of the flare studied. A small fraction of electrons could be accelerated to higher energies ($\sim 1$~MeV), as evidenced by the presence of detectable non-thermal gyrosynchrotron microwave radiation, but could not be registered in the HXR range because of the background. Thus, the lower limit for the maximal kinetic energy of non-thermal electrons is considered here as $K_{0} \sim 0.1$~MeV. From tracing of the flare ribbons we found the magnetic reconnection rate related to the electric field in the current sheet. Firstly, we notice that the reconnection rate estimated for three sub-flares is in qualitative accordance with the observed heating rate (the time derivative of the SXR light curve) and microwave light curves related to the gyrosynchrotron radiation of accelerated electrons. In other words, temporal dynamics of electron acceleration and heating rate correlates roughly with the electric field strength (the reconnection rate) dynamics in the current sheet.

Below we will discuss a possible acceleration process during the first sub-flare when the population of non-thermal electrons was the most energetic among three sub-flares. In the second and third sub-flares, the situation could, in general, repeat the situation in the first sub-flare, but with lower intensity.

The maximum value of the electric field in the current sheet was estimated above in order of magnitude as $E \sim 1/L_{[Mm]} \sim 1$~statvolt/cm (or $\sim 3 \times 10^{2}$~V/cm) for $L=1$~Mm. This electric field highly exceeds the Dreicer field $E_{D} \approx 10^{-11} \ln \Lambda n_{e} T_{e}^{-1} \sim 10^{-6}$~statvolt/cm (or $\sim 10^{-4}$~V/cm), where $\ln \Lambda \approx 20$ is the Coulomb logarithm, $n_{e} \approx 10^{11}$~cm$^{-3}$ and $T_{e} \approx 40$~MK is the electron plasma density and temperature in the acceleration site, respectively (see Paper~I). Considering the acceleration length scale $L_{acc} \approx 10$~Mm, corresponding to the length of the accelerated electron capture region obtained in Paper~I (i.e. the region where the accelerated electrons were trapped and emitted the gyrosynchrotron radiation detected), one can deduce the maximal accelerated electron energy $W_{1} = eEL_{acc} \approx 300$~GeV, which was actually not observed ($W_{1} \gg K_{0}$). This indicates that, most probably, the electron acceleration length scale was several orders of magnitude less than the electron capture length scale. Considering the results of \citet{Litvinenko1996}, one can deduce the maximal energy gained as $W_{2} = e E_{||} \delta B_{||}/B_{\perp}$. Here the current sheet is assumed to be non-neutralized with the longitudinal $B_{||}$ and transverse $B_{\perp}$ magnetic components. For $\delta \sim 6\times 10^4 - 6\times 10^6$~cm (see above) and $B_{||}/B_{\perp} \sim 10$ (that is reasonable for the flare studied, see Paper~I), we obtain $W_{2} \approx 0.2-20$~GeV.

The estimated energies $W_{1}$ and $W_{2}$ of non-thermal electrons are much higher than the observed ones, i.e. $W_1\gg W_2\gg K_0$. It means that if the mechanism of the super-Dreicer electric DC field acceleration is valid, than: (1) the inflow plasma velocity should be $10^{3}-10^{5}$ times lower (i.e. $v_{in}^{\prime} \sim 3.5-350$~m/s), or (2) the reconnection current sheet width ($\delta_{rec}$) should be at least $\sim 10^{3}$ times smaller than the minimum value of the current sheet width $\delta_{min} \approx 6\times 10^4$~cm estimated above from the continuity equation, i.e. $\delta_{rec} \sim 10-100$~cm. The first possibility seems to be unrealistic, as the inflow velocity was estimated from the reconnection rate, which is consistent in order of magnitude with the previous estimations. The second possibility is more likely. Our estimation of $\delta$ is based on the continuity equation~(\ref{eqcont}), where all parameters ($n_{in}$, $n_{0}$, $v_{in}$, $v_{A}$, $l$) can not be measured precisely. Probably, the greatest uncertainty is in estimating the height of the reconnecting current sheet $l$. We took an estimate of $l \sim 1$~Mm from the EUV observation of the bright structure elongated over the PIL. In fact, the height of the current sheet could be much lower, and the apparent thickness of this EUV structure could be determined by the expansion (outflow) of the heated plasma leaving the reconnection region. It is possible that the actual height scale of the reconnecting current sheet may be several orders of magnitude lower than our estimation of $l$, and then the reconnecting current sheet width could be $\delta_{rec} \sim 10-100$~cm. Here we can also add that this $\delta_{rec}$ should be considered as a characteristic spatial scale of electron acceleration to the observed energies. This means that the current sheet itself could be much thicker (with $\delta \sim 6\times 10^4 - 6\times 10^6$~cm), but it had filamentation with a characteristic scale $\delta_{rec}$, which could impede effective acceleration to higher energies.

Let's estimate the electron acceleration time: $\Delta t_{acc} = \sqrt{(B_{||}/B_{\perp})(2\delta_{rec}m_{e})/(eE_{||})} \sim 10^{-8}-10^{-7}$~s, where $m_{e}$ and $e$ is the electron mass and charge, respectively \citep[see ][]{Aschwanden2004}. The resulting estimate shows that electrons are gaining energy very quickly, much faster than the observed time scales. From this point of view, this acceleration process does not contradict the available observations. Electrons can gain energy quickly, and then they can precipitate to the chromosphere or can be trapped in magnetic loops for quite a long time, emitting HXR and microwave radiations. Each observed burst of HXR and microwave emission with time scale of several seconds (or tens of seconds) could consist of millions of ``elementary'' bursts \citep[e.g.][]{Kaufmann1985,Emslie1995}, each of which is associated with acceleration in the current filament of scale $\delta_{rec}$.

Let's also compare the rate of energy gain due to acceleration by the super-Dreicer field ($dW_{SD}/dt$) and the loss rate due to gyrosynchrotron radiation ($dW_{gs}/dt$): $dW_{SD}/dt \sim 10^{7} E_{|| [statvolt/cm]} \sqrt{1-(m_{e}c^{2}/W)^{2}}$~MeV/s and $dW_{gs}/dt \sim -10^{-9} (B_{[G]})^{2} (W/m_{e}c^{2})^{2}$~MeV/s \citep[e.g.][]{Longair1981}. Consequently, $dW_{SD}/dt > \left| dW_{gs}/dt \right|$ for electron energy $W \lesssim 500$~GeV under the estimated $E_{||} \approx 1$~statvolt/cm and $B \approx 1000$~G. This means that for the kinetic energies of accelerated electrons ($K \lesssim 1$~MeV) observed in the flare studied the energy gain due to acceleration in the super-Dreicer field estimated far surpasses the energy loss due to gyrosynchrotron radiation, and the latter can be neglected. The characteristic time of energy loss due to gyrosynchrotron radiation (after leaving the acceleration region) is $\Delta t_{gs} \approx 10^{9} \times (m_{e}c^{2}) K_{0}/[B_{[G]}^{2}(K_{0}+m_{e}c^{2})] \approx 80-330$~s for $K_{0} \sim 0.1-1$~MeV and $B \approx 1000$~G, which is comparable with the duration of the sub-flares observed.

To sum up the above estimations, there is no problem to produce electrons with energies sufficient to generate HXR and microwave emissions by the estimated super-Dreicer electric field. The problem is that such ``ideal'' accelerator seems to be too efficient and could accelerate electrons to much higher energies than the observed ones. Possibly, an efficiency of a real accelerator may be much different due to presence of electric current filamentation (fragmentation). In particular, it was shown on the base of 3D kinetic simulation that electrons can effectively gain energy by a Fermi-like mechanism due to reflection from contracting field lines during reconnection in a filamenting current sheet with a guide magnetic field \citep{Dahlin2015}. A similar situation, but in a much larger physical volume than the simulated one, could well be realized in the flare studied. Another possibility of plasma heating and electron acceleration is related to collapsing magnetic loops (traps), which could have been formed above the PIL as a result of the TCMR (see discussion above). In this process, the kinetic energy gain by a particle is roughly proportional to the square of the ratio of lengths of stretched and unstretched trap \citep{Somov2003,Borissov2016}. It is unlikely that this ratio exceeded $\approx 2-3$ in the confined flare studied. Thus, we suppose that this mechanism, although it could serve as a source of some additional plasma heating and electron acceleration, did not play a crucial role.

From Paper~I we know that the ratio of nonthermal electrons density to thermal super-hot electrons density was $\approx 0.01$. Large-scale electric field acts equally on all electrons, whereas we see that only a small part of them was accelerated. Runaway electrons can quickly setup a strong charge separation that screens the electric field. This process can be also accompanied by excitation of waves/turbulence by beams of runaway electrons and subsequent interaction with them, preventing further effective acceleration \citep[e.g.][]{Boris1970,Holman1985}. On the other hand, it is known that stochastic acceleration, based on wave-particle interaction, is selective to electrons which are in resonance with waves. Thus, only a fraction of electrons from thermal or initially pre-accelerated populations has a chance to be accelerated further. In the above subsection we discussed necessity to introduce filamentation/turbulence to explain tearing instability and the presence of super-hot plasma. Possibly, this turbulence could accelerate electrons in the frame of the stochastic acceleration models. It could explain the ratio of accelerated electrons to thermal particles and less energies than the estimated ones in the frame of the super-Dreicer DC electric field acceleration. Unfortunately, it is difficult to prove the presence of turbulence in the flare studied on the base of the available observational data. Finally, it should be mentioned that other acceleration mechanisms could be also considered \citep[e.g.][and references therein]{Aschwanden2004,Zharkova2011} that is, however, out of the scope of this work.

%---------------------------------------------------------------------------------------------------------------------------------------------------------------------
\subsection{SUMMARY AND CONCLUSIONS}

This work and Paper~I present detailed investigation of the flare energy release in conditions far from those ones assumed in the ``standard'' 2D model of eruptive solar flares. Our interest to the selected confined flare (SOL2015-03-15T22:43), composed of three subsequent sub-flares, was due to its initial energy release development very low in the corona ($H \approx 3$~Mm, possibly even in the chromosphere), in the region with strong vertical PVEC near the the PIL, where we found interaction of the highly stressed sheared magnetic loops. In other words one can say that the selected flare is nice natural experiment showing energy release of the pure three dimensional magnetic reconnection in the confined region (without eruption).

%The observational results obtained in these two papers provided us with the information to discuss the basic properties of the energy release at the PIL for the considered solar flare.

In Paper~I we studied flare in the context of plasma heating and dynamics of nonthermal electrons using multiwavelength observations. In this work we focused on analysis of high-cadence 135-second HMI vector magnetograms that allowed us to investigate dynamics of the photospheric magnetic field and PVEC on flare time scale. We compared maps of photospheric magnetic field components and PVEC with the flare images in different wavelength ranges (optical, UV, EUV, SXR and HXR). Locations of the flare emission sources, as well as the retrieved dynamics of the photospheric magnetic field and extrapolated coronal magnetic field are nicely explained by the TCMR-based flare scenario.

% Analysis revealed that the dynamics of the magnetic reconnection rate (time derivative of the photospheric magnetic flux in the flare ribbons) correlates with heating rate (GOES time derivative) of the separate subflares.

Analysis of 135-second vector magnetograms revealed that the total PVEC in the flare region shows sharp increase during the flare that confirms the observations obtained with 720-second cadence in the previous works. However, we found that the temporal profile of the effective PVEC density (averaged over the flare region around the PIL) has maximum during the first subflare and subsequent gradual decrease. We think that, we found manifestations of electric current reorganization and dissipation connected with the flare energy release process.

From our point of view, the most important result of this paper is the deduced dynamics of the magnetic field and PVEC inside the flare ribbons. High-cadence magnetograms allowed to investigate non-stationary dynamics of the UV ribbons relative to the structure of magnetic field and PVEC in the flare region. For three consecutive sub-flares, we found rough matching of the plasma heating and electron acceleration efficiency with the magnetic reconnection rate, total PVEC and PVEC density in the flare ribbons. We argued that this observation can be also qualitatively interpreted within the TCMR scenario.

Magnetic reconnection (TCMR) was developped in the non-neutralized current sheet (formed between two crossed magnetic structures at the PIL) with dominating guide field $\sim 1000$~Gauss, existed long before the flare onset. Considering the maximal value of magnetic reconnection rate $7\times 10^{18}$~Mx$/$s we found its dimensionless rate $M_A\approx 0.1$ and estimated plasma inflow (into reconnecting current sheet) velocity as $v_{in}\approx 700$~km/s. Reconnection rate is comparable with the upper limit found in other works, where eruptive events were considered. However inflow velocity is larger comparing with the eruptive flares. We think that such fast inflow is not connected with external driver (gradual displacement of magnetic loops due to photspheric motions, or eruption), but is a result of spontaneus equilibrium loss. Estimations also reveal that magnetic reconnection was collisionless and accompanied by formation of thin channels (possibly 3D magnetic islands, see Paper~I) where accelerated electrons and hot plasma were localized.

We would like to emphasize that we hope to continue studies of similar events to extract more detailed quantitative information about flare energy release processes around the PIL, where the three dimensional magnetic reconnection develops. In particular it is still not clear the reason for observed dynamics of the electric currents. To explain it one need data-driven MHD simulations to interprete observations. We also should focus on comparison between magnetic field dynamics and acceleration process with plasma heating. However, to achieve progress in such comparison we should have vector magnitograms with time cadence of one order better than the avalaible ones (135 seconds). Without new generation instruments we can only investigate magnetic reconnection at the PIL indirectly using different assumptions.
	
In conclusion, it is also worth noting that today we have a lot of models of the eruptive flares, but we have a lack of models explaining three dimensional magnetic reconnection (in particular, TCMR) in the conditions similar to the flare studied in this paper. Observational results in our two papers can be used as a input data and expected output for future solar flare models describing three dimensional magnetic restructuring at the PIL. We think that it is important to develop the new self-consisting models of flare energy release reproducing 3D magnetic reconnection in the PIL with strong magnetic field, spatial filamentation of energy release, formation of high energy density populations of nonthermal electrons and appearance of the super-hot plasma.
 
%to extract more useful information about the flare energy release processes around the PIL, where the magnetic reconnection develops in the conditions far from the 2D picture presented in the ''standard`` eruptive flare model, one needs in additional systematic study of the events similar to the discussed ones. 

%and advanced modeling of plasma heating and particle acceleration within the TCMR scenario in realistic systems of highly sheared magnetic loops with the strong guide magnetic field and electric currents.

\acknowledgements

We are grateful to the teams of HMI/SDO, AIA/SDO, RHESSI, SOT/Hinode, NoRP and GOES for the available data used. We thank Drs A.V.~Artemyev and D.Y.~Kolotkov for fruitful discussions. We also appreciate to the anonymous reviewer for a number of useful comments, which helped to improve the paper. This work is supported by the Russian Science Foundation under grant No. 17-72-20134.

%%%%%%%%%%%%%%%%%%%%%%%%%%%%%%%%%%%%%%%%%%%%%%%%%%%%%%%%%%%%%%%%%%%%%%%%%%%%%%%%%%%%%%%%%%%%%%%%%%%%%%%%%%%%%%%%%%%%%%%%%%%%%%%%%%%%%%%%%%%%%%%%%%%%%%%%%%%%%%%%%%%%%%

\bibliographystyle{aasjournal}
%\bibliography{bibl}

\clearpage
%%%%%%%%%%%%%%%%%%%%%%%%%%%%%%%%%%%%%%%%%%%%%%%---------Figures-----------%%%%%%%%%%%%%%%%%%%%%%%%%%%%%%%%%%%%%%%%%%%%%

%_____________________________________OBSERVATIONS__________________________________________

%----------------- time profiles of RHESSI and Nobeyama data ---------------------------------

\begin{figure}[t]
\centering
\includegraphics[width=1.0\linewidth]{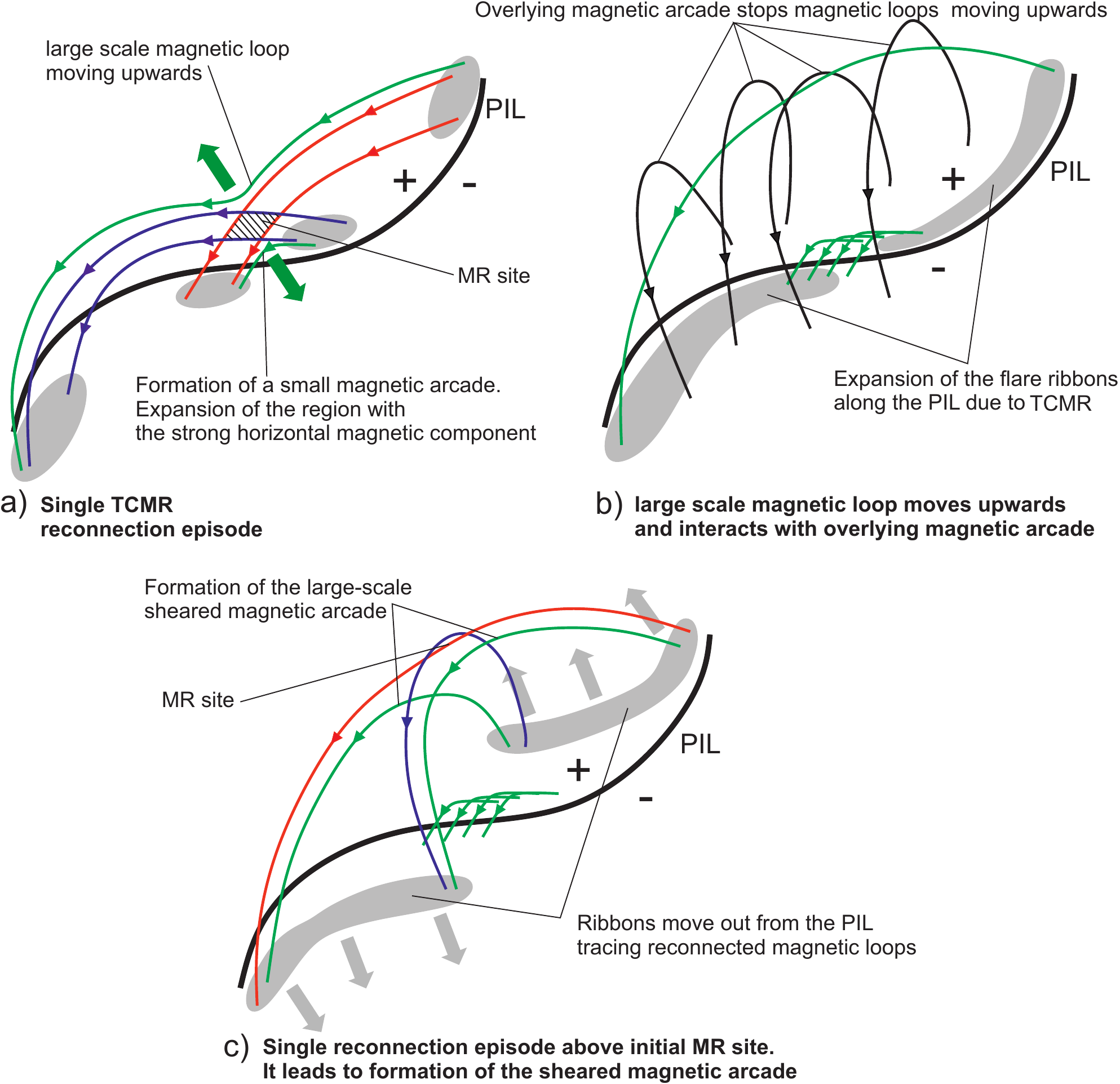}
\caption{This scheme shows our explanations of the flare energy release process. Panel (a) demonstrates the flare onset as TCMR. Then upward moving magnetic field lines interact with the overlaying magnetic arcade (panel~(b)) that leads to the secondary magnetic reconnection (panel~(c)) and cause the flare ribbons to move from the PIL.}
\label{scheme2}
\end{figure}

\clearpage
\end{document}